\documentclass[aps,pre,twocolumn,preprintnumbers,amsmath,amssymb]{revtex4} 
\usepackage{graphicx}
\usepackage{dcolumn}
\usepackage{bm}
\usepackage{tabularx}
\renewcommand{\vec}[1]{{\boldsymbol{#1}}}
\usepackage{subfigure}
\def\eps{\varepsilon}

\def\note#1{\relax}

\begin{document}
\title{The Physics of Traffic and Regional Development}

\author{Dirk Helbing}
  \email{helbing@trafficforum.org}
  \homepage{http://www.helbing.org}
\affiliation{Institute for Economics and Traffic, Dresden University of Technology,
  Andreas-Schubert-Str. 23, 01062 Dresden, Germany
}

\author{Kai Nagel}
  \email{nagel@kainagel.org}
  \homepage{www.kainagel.org}
\affiliation{Transport Systems Analysis and Transport Telematics,
Technical University of Berlin, Salzufer 17--19 SG 12, 10587 Berlin, Germany}

\date{\today}

\begin{abstract}
This contribution summarizes and explains various principles from
physics which 
are used for the simulation of traffic flows in
large street networks, the modeling of destination, transport mode, and route choice,
or the simulation of urban growth and regional development. The
methods stem from many-particle physics, from kinetic gas theory,
or fluiddynamics. They involve energy and entropy
considerations, transfer the law of gravity, apply 
cellular automata and require methods from evolutionary game theory. 
In this way, one can determine interaction forces among driver-vehicle
units, reproduce breakdowns of traffic including
features of synchronized congested flow, or understand 
changing usage patterns of alternative roads. One can also describe 
daily activity patterns based on decision models, simulate migration streams, 
and model urban growth as a particular kind of aggregation process. 
\end{abstract}
\maketitle

\section{Introduction} 
\label{sec:introduction}

\note{multi particle system}

From the point of view of a physicist, cars on a highway or
pedestrians in a mall are many particle systems.  Yet, the particles
of those systems are more complicated than typical particles in
physical systems.  A question therefore is in how far the methods from
physics can be applied to systems with these more complicated
particles.

\note{dynamical states of traffic}

It turns out that in fact quite a lot can be explained with the
methods from physics, which combine simple microscopic modeling
assumptions with advanced mathematics and/or advanced computational
science.  For example, those systems display a surprising variety of
states, such as jammed, laminar, or oscillating.  Those states can be
explained using several concepts, from microscopic to fluid-dynamical.
Physics methods have contributed significantly to better understanding
those states, and to alleviate them where possible.

\note{networks, game theory}

Traffic usually unfolds on \emph{networks}, rather than on flat
two-dimensional space.  And indeed, traffic has some similarity to the steady-state
flow of electrons in a fuse network with non-linear resistors.
Besides the fact that the traffic dynamics can be more complicated
than this steady-state behavior, as pointed out in the last paragraph,
the main difference between human travelers and electrons in a network
is that human travelers typically have an individual destination.  This
introduces an important additional non-linearity, and also brings game
theory into the picture, in particular the concept of a Nash
Equilibrium (NE).  Game theory states that a system \emph{is} at a NE
when no player can gain by unilaterally switching strategies.  What is
more of an interest to physicists is that certain types of learning
systems can have a NE as fixed point attractor -- they can, however,
also have other types of attractors, such as periodic or chaotic ones.
And indeed, route choice experiments with human subjects confirm that
in many cases the players approach a NE, but also that this NE does
not have to be stable and displays behavior reminiscent of
intermittency and volatility clustering.

\note{route choice $\to$ act gen}

This and other evidence demonstrates that the route choice behavior of real
world people can be approximated by a fastest path algorithm, which is
relatively cheap in terms of computational complexity.  From a
practical point of view, the next problem then becomes to generate the
destinations of the travelers.  Two different approaches are discussed
in the paper: (i)~gravity models, which make the probability to select
a destination dependent on the distance; and (ii)~activity-based
demand generation, which attempts to generate complete daily plans for
all travelers in a system.  And once more the situation is similar to
physics: While it may be hopeless to realistically describe the
behavior of individual people in the system, there is some hope that
the macroscopic (``emergent'') behavior of many travelers together can
be derived from simple microscopic rules, akin to the derivation of
the ideal gas equation.  Also, behavioral invariances such as the
energy consumption of the human -- not the vehicle -- can sometimes be
found and used.

\note{city growth}

Intuitively, the next important question is then how the city itself
develops -- where humans find residences, where retail and other
industries find locations, and where the city grows when the space
inside its current boundaries is no longer sufficient.  Once more,
models from physics, this time growth models based on the cellular
automata paradigm, can offer important insight.  The use of such
models is related to the fact that some quantities of real-world
cities, such as the area, the perimeter, or the size distribtion,
display fractal properties.  An alternative approach is to use models
from Synergetics, with its associated master equation, as the starting
point to describe migration dynamics.

Similar to the molecular dynamics approach in physics, it is now
possible to build large scale ``agent-based'' simulation systems of
real world transport systems which are based on the ``first
principles'' introduced above.  Since a typical metropolitan area has
about $10^7$ inhabitants, this is well within the range of
computational feasibility, although some computational efficiency is
lost because of the more complicated particles.  The main challenge
currently is to integrate the different levels, once more a challenge
that is similar to multi-scale modeling in physics.

Results achieved so far indicate that simulation models based on those
principles are, albeit just at the beginning of their development, at
least as good as established methods.  In consequence, one can start
to move those models into the realm of real-world infrastructure
planning.  In addition, one would expect that additional scientific
insights, better implementation techniques allowing faster turnaround,
and the increasingly better availability of electronic data will make
them considerably better in the future.

\section{Fluid-Dynamic and Gas-Kinetic Traffic Models}
\label{sec:fluid-dynamic-gas}

Every driver has probably encountered the widespread phenomenon of so-called ``phantom
traffic jams'', for which there is no visible reason such as an accident or
a bottleneck \cite{Review}. So, why are vehicles sometimes stopped although everyone likes to
drive fast? \cite{Ency} 
\par
The first well-known approach to describe density waves in traffic
flows goes back to Lighthill and Whitham \cite{LiWi} in 1955.  They
start with the typical {\em continuity equation,}
\begin{equation}
 \frac{\partial \rho(x,t)}{\partial t} + \frac{\partial
 Q(x,t)}{\partial x} = 0 \, ,
\label{eq:1}
\end{equation}
where $\rho(x,t)$ denotes density and $Q(x,t)$ denotes flow or
throughput.  The equation reflects that vehicles are not generated or
lost in the absence of ramps, intersections, or accidents. As usual,
one also has the relation $Q(x,t) = \rho(x,t) \, V(x,t)$, where
$V(x,t)$ is the velocity at place $x$ and time $t$. 

Lighthill and Whitham postulated that the traffic flow $Q(x,t)$
could be specified as a function of the density $\rho(x,t)$ only.
This assumes an \emph{instantaneous} adaptation of $Q(x,t)$ to some
equilibrium flow-density relation $Q_{\rm e}(\rho(x,t))$. The
corresponding curve $Q_{\rm e}(\rho) = \rho V_{\rm e}(\rho)$ is often
called the {\em ``fundamental diagram''} and obtained as a fit to
empirical data. 

Because of $\partial Q_e(\rho)/\partial x = ( d Q_e/d\rho )
\,  \partial \rho/\partial x$, one obtains
\begin{equation}
\frac{\partial \rho}{\partial t} + C(\rho) \frac{\partial
\rho}{\partial x} = 0 \ ,
\label{eq:2}
\end{equation}
which is a non-linear wave equation where the propagation velocity
\begin{equation}
C(\rho) = \frac{dQ_{\rm e}(\rho)}{d\rho} = V_{\rm e}(\rho) + \rho \frac{dV_{\rm e}(\rho)}{d\rho} \le V_{\rm e}(\rho)
\label{eq:3}
\end{equation} 
of so-called {\em kinematic waves} depends on the vehicle density
only.  Kinematic waves have the property that they keep their
amplitude, while their shape changes until {\em shock waves} (i.e.
discontinuous changes in the density) have developed.  The densities
$\rho_+$ and $\rho_-$ immediately upstream and downstream of a shock
front determine its propagation speed
\begin{equation}
S(\rho_+,\rho_-) = \frac{Q_{\rm e}(\rho_+)-Q_{\rm
e}(\rho_-)}{\rho_+-\rho_-} = \frac{\Delta Q_{\rm e}}{\Delta \rho} \, .
\label{eq:4}
\end{equation}
Note that Eq.~(\ref{eq:4}) is just the discrete version of Eq.~(\ref{eq:3}).
\par
\par
\begin{figure}[htbp]
  \begin{center}
\includegraphics[width=8cm]{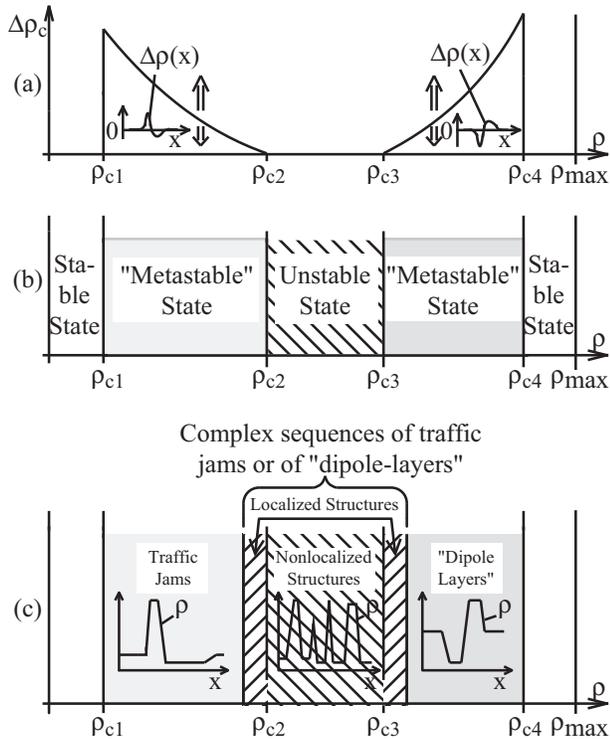}
\end{center}
\caption[]{%
Schematic illustration of (a) the chosen initial perturbation $\Delta
\rho(x)$ and the density-dependent perturbation amplitudes $\Delta
\rho_{\rm c}$ required for jam formation in the meta-stable density
regime, (b) the related instability diagram of homogeneous 
(and slightly inhomogeneous) 
traffic flow (predicting stable traffic at
small and high densities, linearly unstable traffic at medium
densities, and meta-stable traffic in between) and (c) the finally
resulting traffic patterns $\rho(x)$ depending on the respective
density regime.  (Simplified diagram after
\cite{KerKoSchi95,KerKoSc96}, see also \cite{KerKon94}.)
\label{KERNER}}
\end{figure}

Experimental observations of traffic patterns show additional features
which cannot be reproduced by  the above traffic model. While traffic flow appears to be
stable with respect to perturbations at small and large densities, there is a
linearly unstable range $[\rho_{\rm c2},\rho_{\rm c3}]$
at medium densities, where already small disturbances of
uniform traffic flow give rise to traffic jams. Between the stable and linearly unstable
density ranges, one finds meta- or {\em multi-stable} ranges $[\rho_{\rm c1},\rho_{\rm c2})$
and $(\rho_{\rm c3},\rho_{\rm c4}]$, since there exists 
a density-dependent,  critical amplitude $\Delta \rho_{\rm c}$, so that the resulting
traffic pattern is path- or {\em history-dependent} \cite{KerKon94,KerKoSchi95,KerKoSc96,Ker97} 
(see Fig.~\ref{KERNER}). While subcritical perturbations fade away, supercritical perturbations
cause a breakdown of traffic flow (``nucleation effect'').  
Consequently, traffic displays {\em critical points,  
non-equilibrium phase transitions, noise-induced transitions, and fluctuation-induced
ordering phenomena}. One may view this as non-equilibrium analogue of the phase
transitions between vapour (free flow), water (``synchronized'', queued traffic flow above freeway capacity),
and ice (jams with standing vehicles and zero flow).  
Note that noise-like effects enter the above deterministic models
only via the boundary conditions, but some models take care of them by
additional fluctuation terms or by distinguishing different
driver-vehicle types, i.e. partial flows with non-linear interactions.
\par
According to Krau{\ss} \cite{Krauss}, traffic models show the observed {\em hysteretic phase transition}
related with meta-stable traffic and high flows, if the typical maximal acceleration 
is not too large and the deceleration strength is moderate. 
In such kinds of models, $\rho_{\rm c1}$ is the density where the flow-density relation
\begin{equation}
 J(\rho) = \frac{1}{T} \left( 1 - \frac{\rho}{\rho_{\rm jam}} \right) 
\label{jot}
\end{equation}
of traffic with fully developed traffic jams (the so-called
``jam line'') intersects with the free branch of the flow-density
diagram $Q_{\rm e}(\rho)$. $T$ denotes the net time gap (time clearance) in congested
traffic and $\rho_{\rm jam}$ the density inside of traffic jams. 
The intuitive interpretation of Eq.~(\ref{jot}) is that traffic in the
congested regime is composed of moving areas and of areas where
vehicles are stopped.  The fraction of the moving areas is given by $1
- \rho/\rho_{\rm jam}$.

Moreover, the outflow $Q_{\rm out}$ from traffic jams is a
self-organized constant \cite{KerKon94,Ker97,KerReh96a}, which lies
between $Q_{\rm e}(\rho_{\rm c1})$ and $Q_{\rm e}(\rho_{\rm c2})$
\cite{HelHenTre99}.  The jam line corresponds to the flow-density
relation for traffic patterns with a self-organized, stationary
profile \cite{KerKon94}.  These propagate with the velocity $C =
dJ/d\rho = - 1/(T\rho_{\rm jam})\approx -15$~km/h, which is another
traffic constant.  The explanation of this is the
following: Once a traffic jam is fully developed, vehicles leave the
downstream jam front at a constant rate, while new ones join it at the
upstream front.  This makes the jam move upstream with constant
velocity.
\par
In order to reproduce such emergent traffic jams and meta-stable density
regimes, the Lighthill-Whitham model from Eqs.~(\ref{eq:1})
to~(\ref{eq:4}) needs to be generalized. For this, it is helpful that
fluid-dynamic models can be derived from gas-kinetic models, which
relate to models of interactive driver behavior \cite{HelHeShTr01a,HelHeShTr01b}. 
The {\em gas-kinetic approach} has been introduced by Prigogine {\em et al.} \cite{PriAnd,Prig,PriHer}
and is inspired by {\em kinetic gas theory}. It describes the spatio-temporal
change of the phase space density (= vehicle density $\times$ velocity distribution). 
The related equations are either of {\em Boltzmann-like} type (for point-like vehicles or
low densities) or of {\em Enskog-like} type, if vehicular space requirements at moderate and
high densities are taken into account (see \cite{HelHeShTr01a} and references therein). 
They allow the systematic derivation of macroscopic equations for the 
vehicle density $\rho(x,t)$, the average velocity $V(x,t)$, the velocity variance $\Theta(x,t)$, etc. 
This hierarchy of equations is usually closed after the velocity or variance equation,
although the separation of time scales assumed by the underlying
approximations is weak.
Nevertheless, the observed traffic dynamics is rather well reproduced by the resulting 
coupled partial differential equations. The density equation is just the continuity
equation 
\begin{equation}
 \frac{\partial \rho}{\partial t} + \frac{\partial (\rho V)}{\partial x} = 
 \nu_+ - \nu_- \, ,
\end{equation}
where  $\nu_+$ and $\nu_-$ denote on- and off-ramp flows, respectively. The velocity equation can
be cast into the form
\begin{equation}
 \frac{\partial V}{\partial t} + V \frac{\partial V}{\partial x} =
 - \frac{1}{\rho} \frac{\partial P}{\partial x} + \frac{1}{\tau}
 (V^{\rm e} - V) \, .
\label{eq:vel}
\end{equation} 
In theoretically consistent macroscopic traffic models such as the gas-kinetic-based
traffic model, the ``traffic pressure'' $P$ and the velocity $V^{\rm e}$ are non-local
functions of the density $\rho$, the average velocity $V$, and the variance $\Theta$
\cite{HeTre98,ShvHe99}. 
Several well-known models are special cases of Eq.~(\ref{eq:vel}):
\begin{itemize}

\item The Lighthill-Whitham model of Eqs.~(\ref{eq:1}) to~(\ref{eq:4})
is obtained in the (unrealistic) limit $\tau \rightarrow 0$ of
vanishing adaptation times $\tau$.  In that case $V(x,t) =
V_e(\rho(x,t))$, and therefore $Q(x,t) = \rho(x,t) \, V_e(\rho(x,t))
=: Q_e( \rho(x,t))$, which depends on density only as was postulated
for the LW model.

\item
Payne's macroscopic traffic model \cite{Payne1,Payne2} is obtained for
$P(\rho) = [V_{\rm max} - V_{\rm e}(\rho)]/(2\tau)$ and $V^{\rm e} = V_{\rm e}(\rho)$,
where $V_{\rm max}$ denotes the maximum speed (the average velocity at
very low densities). 

\item
Kerner's and Konh\"auser's model \cite{KerKon93,KerKon94} is a variant of K\"uhne's model \cite{Kuehne}
and corresponds to the specifications $P = \rho \Theta_0
- \eta_0 \partial V/\partial x$ and $V^{\rm e} = V_{\rm e}(\rho)$, where $\Theta_0$ and $\eta_0$
are positive constants. The corresponding equation is a Navier-Stokes equation with a
viscosity term $\eta_0 \partial^2 V/\partial x^2$ and an additional relaxation term
$[V_{\rm e}(\rho)-V]/\tau$ describing the delayed adaptation to the velocity-density relation
$V_{\rm e}(\rho)$. 

\end{itemize}
Both for the Payne model and for the Kerner-Konh{\"a}user model, 
the linearly unstable regime $[\rho_{\rm c2},\rho_{\rm c3}]$ 
is determined by the densities $\rho$ fulfilling the condition
\begin{equation}
  \rho \left| \frac{dV_{\rm e}(\rho)}{d\rho} \right|
 > \sqrt{ \frac{dP(\rho)}{d\rho} } \, .
\end{equation}
In consequence, both the Payne model and the Kerner-Konh\"auser model
have linearly unstable ranges if $dV_{\rm e}(\rho)/d\rho$ is large,
while the Lighthill-Whitham model is marginally stable.
\par
Beyond linear stability analysis, in some cases it is also
possible to find analytical expressions for the large amplitude
instability in the meta-stable ranges $[\rho_{\rm c1}, \rho_{\rm c2})$
and $(\rho_{\rm c3},\rho_{\rm c4}]$ (Fig.~\ref{KERNER}).  For these more complicated
aspects, the interested reader is referred to
\cite{Kerner:Konh:large:amplitude}. 
\par
In simple terms, the reason for emergent traffic jams is as follows:
Due to the finite adaptation time (= reaction + acceleration time), 
a small disturbance in the traffic flow $Q_{\rm e}$ can cause an {\em overreaction} (overbraking)
of a driver, if the safe vehicle speed $V_{\rm e}(\rho)$ (which satisfies $dV_{\rm e}(\rho)/d\rho \le 0$) 
drops too rapidly with increasing vehicle density $\rho$. 
At high enough densities $\rho$, this will give rise to a {\em chain reaction} of the followers, 
as other vehicles will have approached before the original speed can be regained.
This {\em feedback} can eventually cause the unexpected standstill of vehicles known
as traffic jam. 

\section{Congested Traffic States in Theory and Empirical Data}
\label{sec:cong-traff-stat}

With this knowledge one can understand the various congested traffic states observed 
on freeway sections with bottlenecks \cite{HelHenTre99,Lee3LeKi99,Lee3LeKi00,inTGF01,TS}.
Given that the flow $Q$ and density $\rho$
are measured {\em per lane,} a bottleneck due to ramp flows is determined by
$\nu_+ = Q_{\rm rmp}/(IL)$, where $L$ is the used length of the on-ramp and $I$ the
number of freeway lanes. The corresponding bottleneck strength is
\begin{equation}
\Delta Q = \frac{Q_{\rm rmp}}{I} \,  . 
\end{equation}

\subsection{Theoretical Phase Diagram of Traffic States}

\begin{figure*}[htbp]
\includegraphics[width=0.75\hsize]{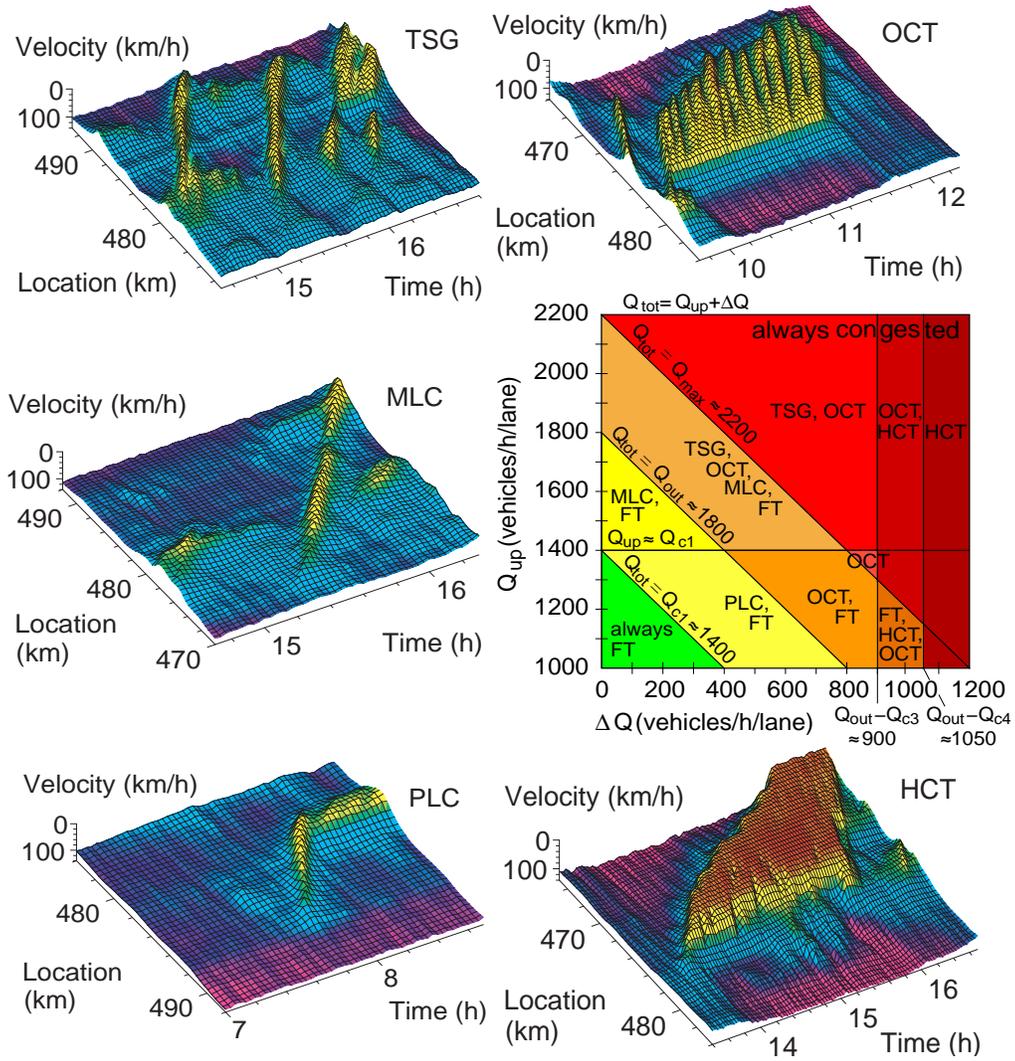}
\caption[]{Empirical representatives
of different kinds of congested traffic observed on the German freeway A5 close
to Frankfurt (after \cite{Ency,TS})
and theoretical phase diagram of traffic states (after \cite{TS}) in
the presence of one bottleneck as a function of the
upstream flow $Q_{\rm up}$ and the bottleneck strength
$\Delta Q$ (center right).}
\label{fig.1}
\end{figure*}
Let $Q_{\rm up}$ denote the traffic flow per lane 
upstream of the bottleneck and
\begin{equation}
Q_{\rm tot} = Q_{\rm up} + \Delta Q = Q_{\rm up} + \frac{Q_{\rm rmp}}{I} 
\end{equation}
the total capacity  required downstream of the ramp. Then, we expect
to always observe free traffic (FT) below the first instability threshold at density $\rho_{\rm c1}$, i.e. if
\begin{equation}
 Q_{\rm tot} = Q_{\rm up} + \Delta Q < Q_{\rm e}(\rho_{\rm c1}) \, .
\end{equation}
Traffic flow will always be congested, if the maximum flow
$Q_{\rm max} = \max_\rho Q_{\rm e}(\rho)$ (the capacity) is exceeded, i.e.
\begin{equation}
 Q_{\rm tot} = Q_{\rm up} + \Delta Q > Q_{\rm max} \, .
\end{equation}
At least if the density related with the maximum flow $Q_{\rm max}$ lies between $\rho_{\rm
  c1}$ and $\rho_{\rm c2}$, traffic states between the two diagonal
lines $Q_{\rm up} + \Delta Q = Q_{\rm e}(\rho_{\rm c1})$ and
$Q_{\rm up} + \Delta Q = Q_{\rm max}$ in the $Q_{\rm up}$-vs-$\Delta Q$
phase space can be either congested or free, depending
on the initial and boundary conditions \cite{TS} (see Fig.~\ref{fig.1}). While homogeneous free flow may persist over
long time periods, large perturbations tend to
produce congested states. Extended congested traffic can emerge above the line 
\begin{equation}
Q_{\rm up} =Q_{\rm out} - \Delta Q 
\end{equation}
in the phase diagram, i.e.
if $Q_{\rm tot}$ is greater than the dynamic capacity $Q_{\rm
  out}$. This line does not have to be parallel to the previously 
mentioned phase boundaries, as $Q_{\rm out}$ may depend 
on the bottleneck strength $\Delta Q$ \cite{TreHeHe00,Review}. 
For $Q_{\rm e}(\rho_{\rm c1}) \le Q_{\rm tot} = Q_{\rm up} +
\Delta Q < Q_{\rm out}$, congested traffic states are always
localized, i.e. they never grow over long sections of the freeway, as the outflow can cope with
the overall traffic volume.
\par
The traffic flow $Q_{\rm cong}$ resulting in the congested
area plus the inflow or bottleneck strength $\Delta Q$ are normally
given by the outflow $Q_{\rm out}$, i.e.
\begin{equation}
 Q_{\rm cong} = Q_{\rm out} - \Delta Q
\end{equation}
(if vehicles cannot enter the freeway downstream of the congestion front).
One can distinguish the following cases:
{\em Homogeneous congested traffic} (HCT) such
as typical traffic jams during holiday seasons can occur,
if the density $\rho_{\rm cong}$ associated with the congested flow 
\begin{equation}
Q_{\rm cong} = Q_{\rm e}(\rho_{\rm cong})
\end{equation} 
lies in the stable or meta-stable range 
\begin{equation}
 Q_{\rm cong} < Q_{\rm e}(\rho_{\rm c3}) \, , \quad \mbox{i.e.} \quad 
 \Delta Q > Q_{\rm out} - Q_{\rm e}(\rho_{\rm c3}) \, .
\end{equation}
Oscillating forms of congested traffic can emerge, if
\begin{equation}
 \Delta Q \le Q_{\rm out} - Q_{\rm e}(\rho_{\rm c4}) \quad \mbox{and} \quad
 Q_{\rm up} > Q_{\rm out} - \Delta Q \, .
\end{equation}
That is, lower bottleneck strengths tend to produce less serious congestion, namely
oscillating rather than homogeneous congested flow. We either find
{\em oscillating congested traffic} (OCT), {\em triggered stop-and-go
  traffic} (TSG), or {\em moving localized clusters} (MLC) \cite{TS}. In contrast to OCT, stop-and-go
traffic is characterized by a sequence of moving jams, between which traffic flows freely.
This state can either emerge from a spatial sequence of homogeneous and oscillating
congested traffic \cite{Koshi}, which is also called the {\em ``pinch effect''} \cite{pinch}, 
or it can be caused by the inhomogeneity at the ramp. In the latter case, 
each traffic jam triggers another one by inducing a small perturbation 
in the inhomogeneous freeway section (see Fig.~\ref{fig.1}), 
which propagates downstream as long as it is small, but turns back when it has grown
large enough {\em (``boomerang effect'')}. This growth requires the downstream traffic flow
to be linearly unstable. If it is meta-stable (when the traffic volume $Q_{\rm tot}$
is sufficiently small), small perturbations will fade away. Therefore, if 
\begin{equation}
 Q_{\rm e}(\rho_{\rm c1}) \le  Q_{\rm tot} = Q_{\rm up} + \Delta Q \le Q_{\rm out} \le Q_{\rm e}(\rho_{\rm c2}) \, ,
\end{equation}
one expects to find localized traffic states, either
a single {\em moving localized cluster} (MLC), or a {\em pinned localized cluster} (PLC)
at the location of the ramp. The latter requires the traffic flow in the upstream section to be stable, i.e.
\begin{equation}
 Q_{\rm up} < Q_{\rm e}(\rho_{\rm c1}) \, , 
\end{equation}
so that no traffic jam can survive there. In contrast, moving
localized clusters and triggered stop-and-go waves require \cite{TS}
\begin{equation}
Q_{\rm up} \ge Q_{\rm e}(\rho_{\rm c1}) \, .
\end{equation}
The simulation results displayed in Figure~\ref{fig.1} summarize, 
for freeways with a single bottleneck, which states are
typically found in different areas of the phase diagram. Apart from
the detailed shape and exact location of the phase boundaries, this phase
diagram is expected to be universal for all
microscopic and macroscopic, stochastic and deterministic traffic models with the 
same {\em instability diagram} (with stable, meta-stable, and unstable
density ranges) \cite{Review}. 
Results for more complex freeway geometries are available as well.

\subsection{Wide Scattering of Traffic Data in Heterogeneous Traffic}

\begin{figure}[htbp]
\begin{center}
\includegraphics[height=4.2cm,angle=-90]{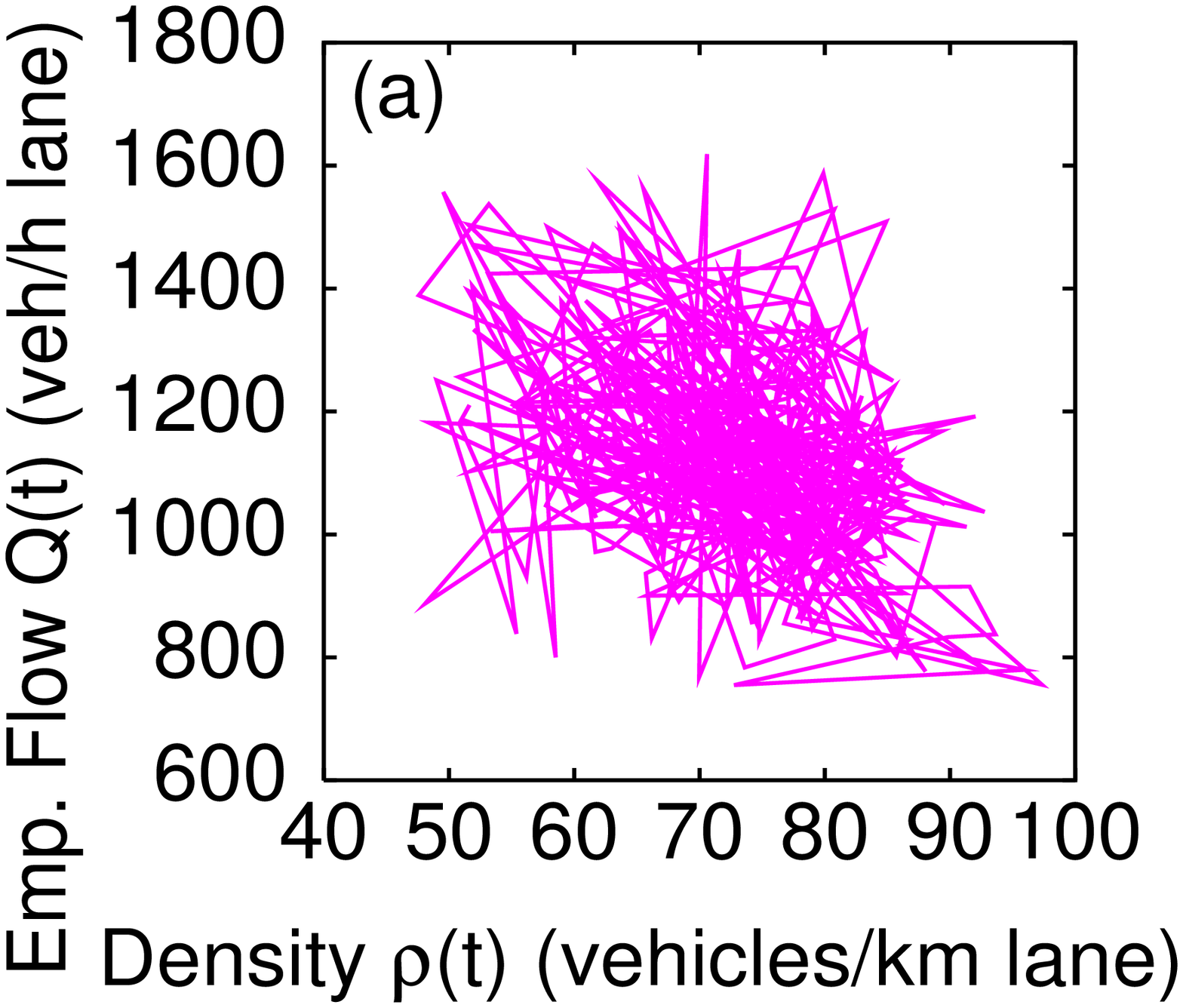}
\includegraphics[height=4.2cm,angle=-90]{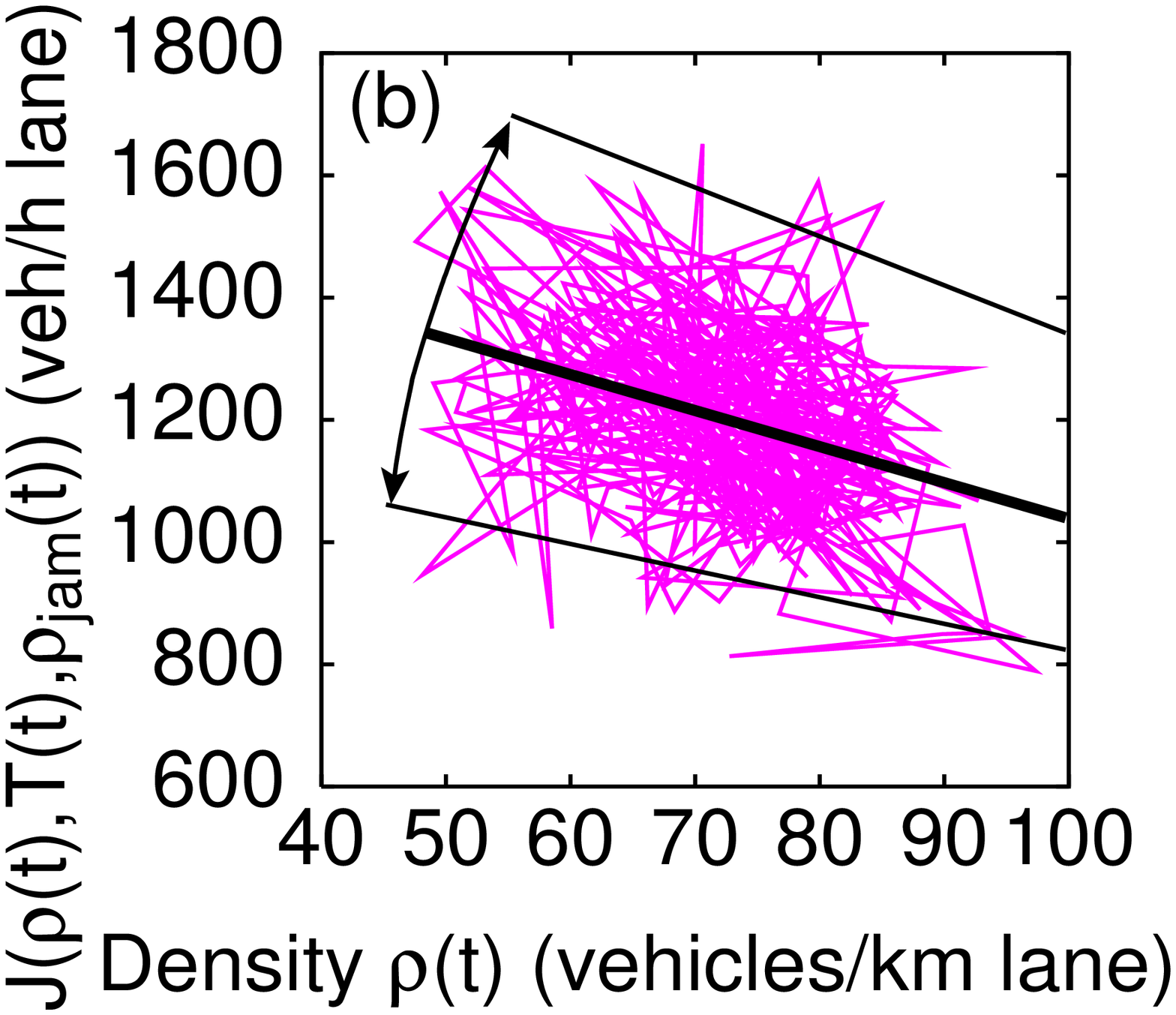}
\end{center}
\caption[]{The two-dimensional scattering of empirical
flow-density data in synchronized traffic flow of high density $\rho \ge 45$ veh/km/lane
(see (a)) is well reproduced by the jam relation (\ref{jot}), when not
only the variation of the density $\rho$, but also the empirically measured variation of
the average time gap $T$ [and the maximum density $\rho_{\rm max}$] is taken into account (see (b)).
The pure density-dependence $J(\rho)$ (thick black line) is linear and 
cannot explain a two-dimensional scattering. However, variations of the
average time gap $T$ change its slope $-1/(\rho_{\rm max}T)$ (see arrows), 
and about 95\% of the data are located 
between the thin lines $J(\rho,T\pm 2\Delta T,1/l) = (1-\rho l)/(T\pm 2 \Delta T)$,
where $l = 3.6$~m is the average vehicle length, $\overline{T}=2.25$~s the average time gap, and
$\Delta T=0.29$~s the standard deviation of $T$.  The predicted form of this area is club-shaped, as
demanded by Kerner \cite{critic1,critic2,critic3}. (After \cite{Kats}.)
\label{synchro}}
\end{figure}
In the congested traffic regime, flow-density data are characterized by a surprisingly
wide scattering \cite{KerReh96b} (see Fig.~\ref{synchro}a). This has led people to question the
applicability of the jam line (\ref{jot}) \cite{critic1,critic2,critic3}. However, it can be shown that, taking into account
the variation of the average net time gap (time clearance) $T$, the variations in the
data can be reproduced with a correlation coefficient of 0.92 (compared to 0.35 when only the density is
varied). 
\par\begin{figure*}[htbp]
\begin{center}
\unitlength1.0cm
\begin{picture}(12,8)
\put(-0.25,3.9){\includegraphics[width=8.3\unitlength]{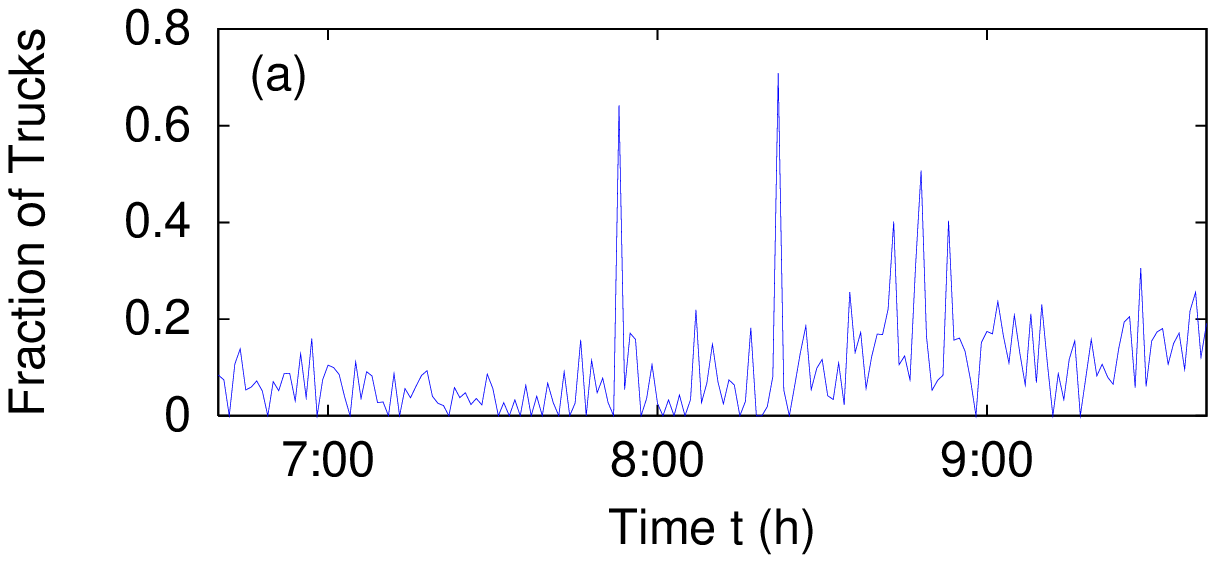}}
\put(7.9,4.3){\includegraphics[width=4.2\unitlength]{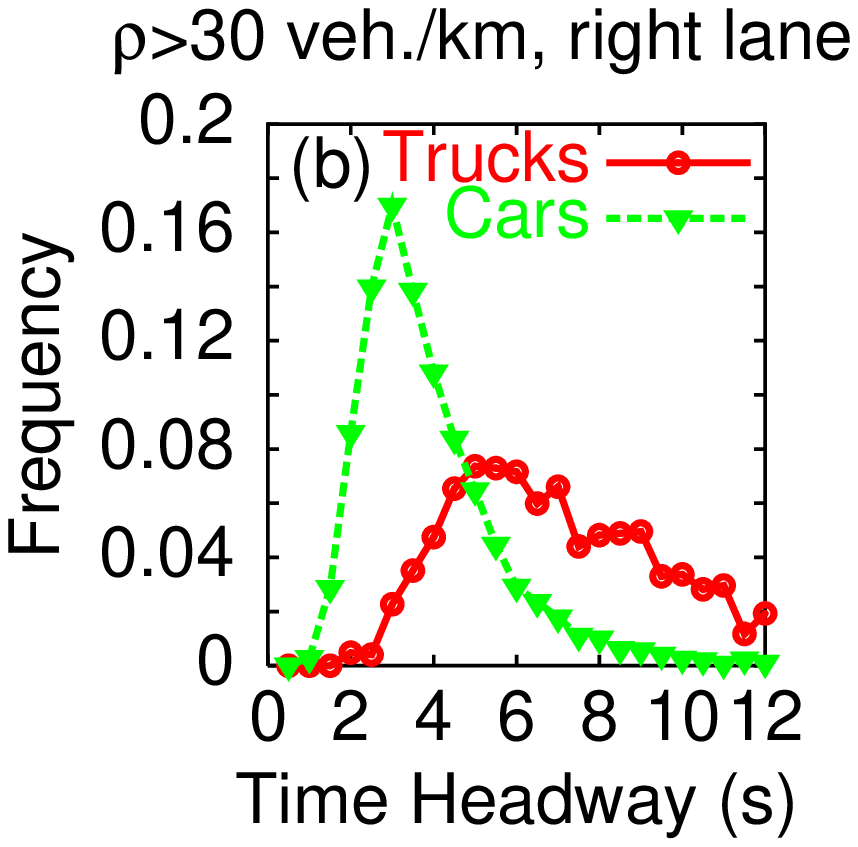}}
\put(-0.5,0.1){\includegraphics[width=4.6\unitlength]{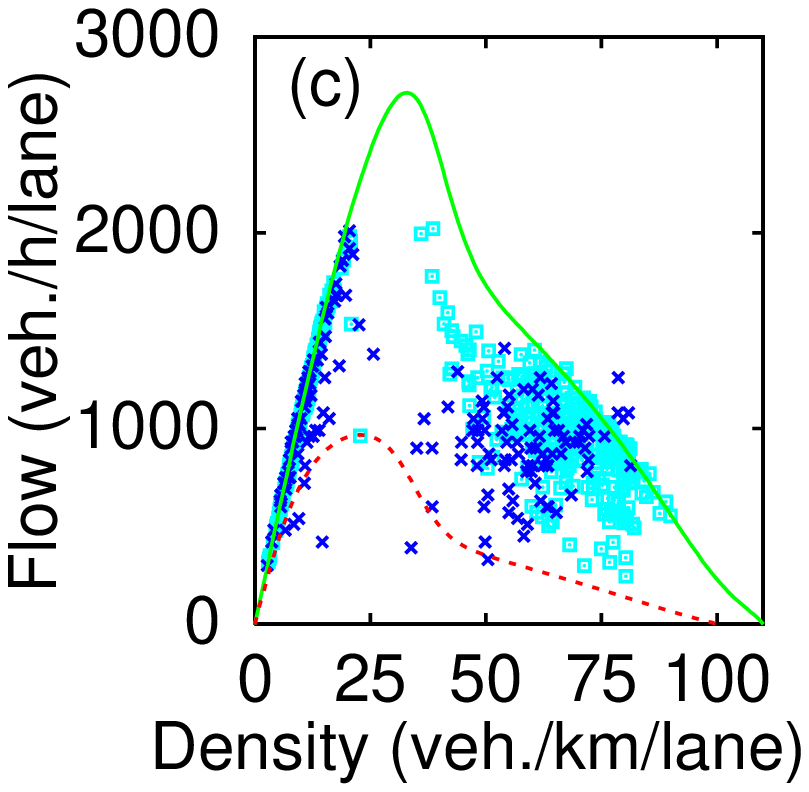}} 
\put(3.5,-0.3){\includegraphics[width=4.52\unitlength]{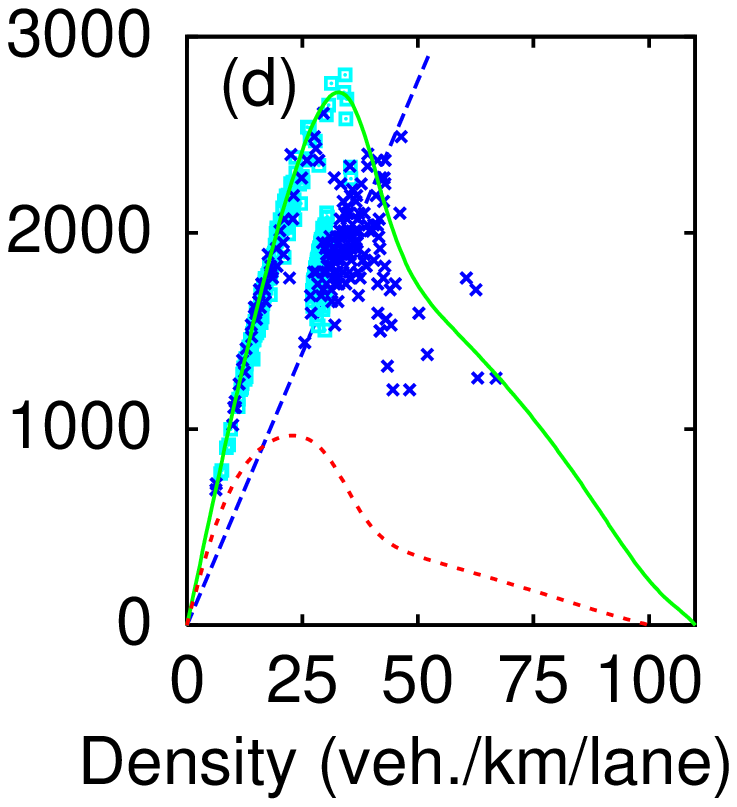}} 
\put(7.7,-0.3){\includegraphics[width=4.2\unitlength]{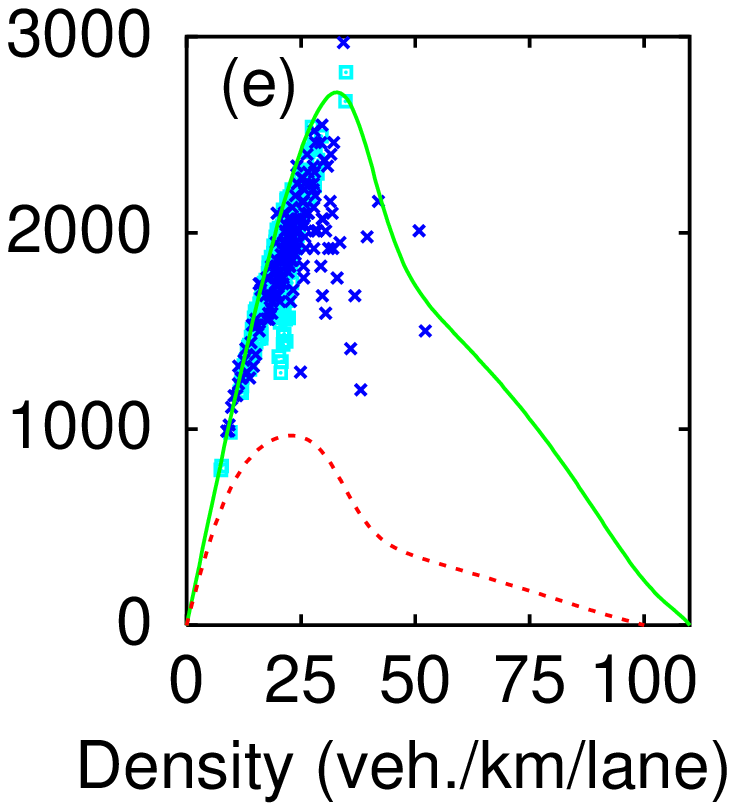}} 
\end{picture}
\end{center}
\caption[]{(a) The empirical truck fraction varies considerably in the course of time.
(b) The time headways of long vehicles (``trucks'') are on average much higher than
those of short vehicles (``cars''). (c)-(e) Assuming a fundamental diagram for cars (solid line), a separate
one for trucks (dashed line), weighting them according to the measured truck fraction, and 
using empirical boundary conditions, allows one to reproduce the observations in a 
(semi-)quantitatively way \cite{Scattering}: (c) Free traffic (at low densities) is characterized by a
(quasi-)one-dimensional curve. Data of congested traffic {\em upstream} of a bottleneck 
are widely scattered in a two-dimensional area. (d) {\em Immediately downstream} of the bottleneck,
one observes homogeneous-in-speed states reflecting recovering traffic. (e) Further
downstream, the data points approach the curve describing free traffic. Dark symbols correspond
to empirical one-minute data, light ones to corresponding simulation results. (After \cite{inTGF01}.)}
\label{wide}
\end{figure*}
A closer analysis reveals a large variation of time gaps between vehicles. The time gaps of trucks
are particularly large. Simulations of heterogeneous traffic with different kinds of vehicles
(i.e. different parameter values) suggest that at least part of the scattering of flow-density data
may be explained by the mixture of cars and trucks \cite{Scattering} and of different driving 
behaviors (see Fig.~\ref{wide}). Although heterogeneous traffic can be treated by macroscopic models (see Fig.~\ref{wide}),
a simulation of individual vehicles is much more easy, flexible, and efficient, in particular if one likes to simulate
network traffic of drivers with different origins and destinations. Therefore, we will focus on
some ``microscopic'' traffic models in the following.

\section{Driven Many-Particle Systems and Microscopic Traffic Simulation}
\label{sec:driven-many-particle}

The observations in freeway traffic can not only be described by fluid-dynamic or
``macroscopic'' traffic models. They are also well reproduced by ``microscopic'' models \cite{TreHeHe00}.
Note that a theoretical connection between both approaches exists. It is called the
{\em micro-macro link} \cite{HelHeShTr01b}. 

\subsection{Car-Following Models}

{\em Microscopic models} are often {\em follow-the-leader 
models} specifying the acceleration $dv_i/dt$ of each
single vehicle $i$ as a function $F_i$ of their speed $v_i$, their 
distance headway $d_i = x_{i-1} - x_i$ with respect to the leading vehicle $i-1$,
and/or their relative velocity $\Delta v_i = v_i - v_{i-1}$:
\begin{equation}
 \frac{dv_i}{dt} = F_i(d_i,v_i,\Delta v_i) + \xi_i(t)\, .
\end{equation}
$\xi_i(t)$ are random variations in the acceleration behavior,
while $F_i$ may be interpreted as average situation-dependent acceleration force.
A typical example is the non-integer car-following model
\begin{equation}
 \frac{dv_i(t+\Delta t)}{dt}
 = - \frac{\Delta v_i(t)}{\tau} \frac{[v_i(t+\Delta t)]^{\ell}}{[d_i(t)]^{m}} 
\end{equation}
with the reaction time $\Delta t \approx 1.3$s and the parameters $\tau \approx \Delta t / 0.55$,
$\ell \approx 0.8$, and $m \approx 2.8$ \cite{Gazis}. It has a linearly unstable range for
$\tau / T > 1/2$. (The time delay $\Delta t$ on both sides facilitates to determine
analytical solutions.) A simpler model is the optimal velocity model
\begin{equation}
 \frac{dv_i(t)}{dt} = \frac{1}{\tau} \Big[ V\big(d_i(t)\big) - v_i(t) \Big] \, ,
\label{Band}
\end{equation}
where $V(d_i)$ is the ``optimal'' velocity-distance relation and $\tau$ the adaptation 
time \cite{Bando1,Bando2}. This model has an unstable range for $dV(d_i)/dd_i > 1/(2\tau)$.
The respective non-linearly coupled differential equations (or stochastic differential
equations, if fluctuations are taken into account) are numerically solved as in molecular dynamics. 
\par
Let us now define the maximum, desired, or free velocity by $V_{\rm max} = \max V(d_i)$. 
The expression 
\begin{equation}
f(s_i) = \frac{V(s_i+l_i) - V_{\rm max}}{\tau} 
\end{equation}
reflects the interaction force $f$ as a function of the net 
distance (clearance) $s_i = d_i - l_i = (x_{i-1} - x_i - l_i)$ between two successive vehicles, where
$l_i$ denotes the length of vehicle $i$. 
With this, Eq.~(\ref{Band}) can be reformulated
in terms of the {\em ``social force model''}
\begin{equation}
 \frac{dv_i}{dt} = \frac{V_{\rm max} - v_i}{\tau} + f(s_i) + \xi_i(t) \, .
\end{equation}

\subsection{Interaction Forces Among Vehicles} 

The difficulty is now to specify the interaction forces among vehicles in an appropriate way.
Their exact specification is essential for realistic traffic simulations,
which are required for the design of efficient and reliable traffic optimization measures such as
intelligent on-ramp controls, driver assistance systems, lane-changing assistants, etc.
If the fluctuation term was zero (i.e. $\xi_i(t) = 0$), the interaction force could simply be
determined from the velocity-density relation. Since the average of the headways $d_i = s_i + l_i$ determines
the inverse density $1/\rho$, we would simply find the relation $f(s_i) = [V_{\rm e}(1/(s_i+l_i)) - V_{\rm max}]/\tau$.
\par
However, things are considerably more difficult, when random fluctuations are not negligible.
One may think of applying scattering theory to determine the interaction potential $U(s) = - df(s)/ds$
from  statistical distributions, but learning about human
interactions requires a somewhat different approach. Progress has recently
been made by means of {\em random matrix theory,} a powerful method
from quantum and statistical physics. It allows one to determine the
interaction potentials via net distance distributions of vehicles.
\par
Random matrix theory \cite{Mehta} has been developed for many-particle 
systems exposed to a "thermal bath" of a given reciprocal
temperature $\beta$, i.e. to random forces of a certain variance
and statistics. The resulting velocity and net distance distributions 
allow one to draw conclusions about the interaction
potential $U(s) = - df(s)/ds$, as this determines their shapes. 
Although a generalization to arbitrary potentials $U(s)$ is possible, 
here we will focus our attention on power-law potentials 
\begin{equation}
U(s) \propto s^{-\alpha} 
\end{equation}
for simplicity. The exponent $\alpha > 0$ is a fit parameter and $s$ is the net 
distance (clearance) between two successive cars. Such a potential describes the
repulsive tendency of drivers to keep a safe distance to the respective car ahead. 
The related net distance distributions can be calculated to be
\begin{equation}
P(\rho s)=Ae^{-\beta(\rho
s)^{-\alpha}}e^{-B\rho s} \, ,
\label{gapdist}
\end{equation}
where $\rho$ is the vehicle density and $A$, $B$ represent 
normalization constants determined by requiring $\int_0^\infty
P(\rho s)\,d(\rho s)=1$ and $\langle \rho s \rangle
\equiv \int_0^\infty \rho s P(\rho s)\,d(\rho
s)=1.$
\par
Originally, the applied random matrix method is an equilibrium concept 
assuming the conservation of energy ${\cal H}={\cal T}+{\cal V}$, i.e. a transformation of
potential energy ${\cal V}=\sum_{i=1}^{N} U(|x_{i-1}-x_i - l_i|)$ into
kinetic energy ${\cal T}=\sum_{i=1}^N m_i v_i^2/2$ with $x_i$, $l_i$, $m_i$, 
and $v_i$ being the location, length, mass, and velocity of the $i$th
vehicle. It can however be shown \cite{fpg} that this is a reasonable approximation for an
ensemble of $N$ vehicles $i$ in a stationary state at
a given density $\rho$ and inverse ``temperature'' (velocity variance) $\beta$, 
while the stop-and-go regime between 20 and 40
vehicles per kilometer and lane must be excluded from an investigation of this kind. 
\par
\begin{figure}[htbp]
\begin{center}
\includegraphics[width=8.5cm, angle=0]{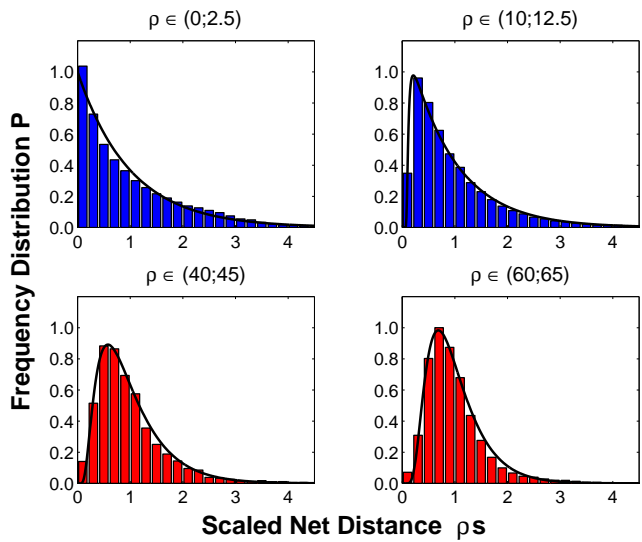}
\end{center}
\caption[]{Representative distributions of net distances (clearances) among 
successive cars for various densities $\rho$, determined from
Dutch single-vehicle data (after \cite{PhysJ}). The solid 
fit curves correspond to the theoretical distributions (\ref{gapdist}) for
the empirical values of $\rho$ and $\beta$. In the free traffic regime,
the only fit parameter is $\alpha = 4$ (above, blue), while $\alpha =1$ in
the congested regime (below, red).}
\label{gap1}
\end{figure}
\begin{figure}[htbp]
\begin{center}
\includegraphics[width=8.5cm, angle=0]{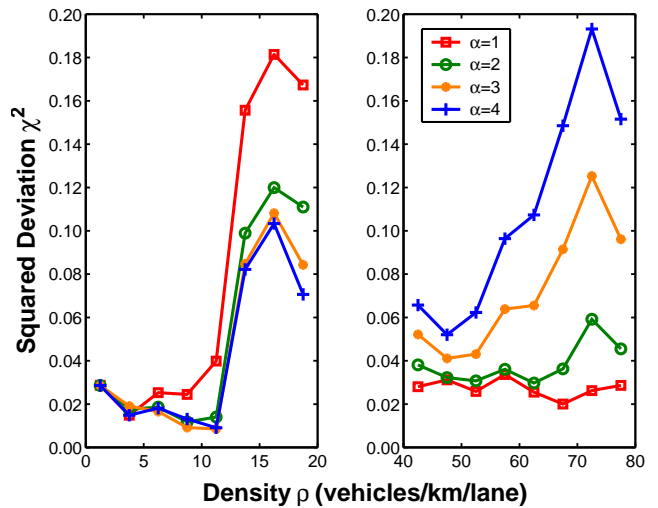}
\end{center}
\caption[]{Sum of squared deviations between the empirical and
theoretical net distance distributions for various fit parameters $\alpha$ (after \cite{PhysJ}).
The best fit is $\alpha = 4$ in the free traffic regime and $\alpha = 1$ in the congested
regime.}
\label{gap2}
\end{figure}

Milan Krbalek and one of us have separately analyzed eight density intervals of Dutch single vehicle data 
in the free low-density regime and eight density intervals
in the congested traffic regime. In accordance with the theoretical predictions,
the velocity distributions fit Gaussian distributions very well \cite{milan}. Moreover,
the best fit of the net distance distributions is obtained for the integer parameter $\alpha = 4$
in the free traffic regime, while we find an excellent agreement with $\alpha = 1$ 
in the congested regime (see Figs.~\ref{gap1}, \ref{gap2}). This is  not only a strong support of
studies questioning a uniform behavior of drivers in all traffic regimes \cite{Neubert}. It also offers
an interpretation of the mysterious fractional distance-scaling exponent $\alpha+1 \approx
2.8$ in  classical follow-the-leader models \cite{MayKe67}, which average over the driver behavior
in the free and congested regimes. 
\par
Note that the net distance distribution (\ref{gapdist}) agrees with the Pearson III type
of distribution \cite{May} which has been suggested to fit empirical
time headway distributions already before our theoretical explanation had been found. Therefore, the
determination of interaction potentials in freeway traffic contributes to the 
challenging problem of understanding time headway or distance distributions of cars \cite{May,Hoog}. 
In addition, 
the identification of behavioral regimes points to adaptive driver behavior. 

\section{Simulation of Large Traffic Networks with Cellular Automata}
\label{sec:simul-large-traff}

\subsection{Cellular Automata Rules for Links}

An alternative approach to car-following models are rule-based 
cellular automata (CA). For the CA technique, road links are divided
into cells, say of length 7.5~m, which can contain at most one car each, and there is one array of cells for
each lane, see Fig.~\ref{fig:gapdef}.  Movement is performed by jumping from one
cell to another, where the new cell is determined by a set of ``driving rules.''  
A good update time step is 1~sec (justified by
reaction time arguments).  Taking this together means, for example,
that a jump of $5$ cells in a time step models a speed of $5 \times
7.5$~m/sec = 135~km/h.

Driving rules of traffic on a link consist of two parts: car following, and
lane changing.  Typical rules for car following are: deterministic
speed calculation, randomization, and movement.
The \textit{deterministic speed calculation} rule first computes a
new speed for each car based on its current speed and closeness to the
car in front of it.  An example for such a rule is
\[
v_{{\rm safe},t} = \min[ v_{t-1}+1, v_{{\rm max}}, g_t] \ ,
\]
where $v_{\rm t-1} +1$ models acceleration, $v_{\rm max}$ is a speed limit, and $g_t$
is the gap, i.e.\ the number of empty spaces to the vehicle ahead.  See Fig.~\ref{fig:gapdef} for an illustration of gap.
Next, the \textit{randomization} rule introduces a probability
$p_{{\it noise}}$ for each car so
that instead of following the above speed deterministically, it may
instead drive one velocity level more slowly:
\[
v'_t = 
\begin{cases}
\max[0,v_{{\rm safe},t}-1] & \hbox{with probability } p_{\rm noise} \cr 
v_{{\rm safe},t} & \mbox{else.} \cr 
\end{cases}
\ .
\]
In the 
\textit{movement} rule, each particle is moved forward:
\[
x_{t+1} = x_t + v'_t \ .
\]
To illustrate the above rules, Fig.~\ref{fig:gapdef} shows traffic moving to
the right.  Using a $p_{\rm noise}$ value of 0.2, the
leftmost vehicle accelerates to velocity~2 with
probability 0.8 and stays at velocity~1 with probability 0.2.  The
middle vehicle slows down to velocity~1 with probability 0.8 and to
velocity~0 with probability 0.2.  The rightmost vehicle accelerates
to velocity~3 with probability 0.8 and stays at velocity~2 with
probability 0.2.  Velocities are in ``cells per time step.''  All
vehicles are then moved according to their velocities using the
movement rule described above.

Fig.~\ref{fig:stca} shows, in a so-called space-time diagram for
traffic, the emergence of a traffic jam.  Lines show configurations of
a segment of road in second-by-second time steps, with time increasing in the downward direction; cars drive from left
to right.  Integer numbers denote the velocities of the cars.  For
example, a vehicle with speed ``3'' will move three sites (dots)
forward.  Via this mechanism, one can follow the movement of vehicles
from left to right, as indicated by an example trajectory.

\subsection{Lane Changing}

\begin{figure}[tp]
\centerline{%
\subfigure[]{%
\begin{minipage}[c]{0.75\hsize}
\includegraphics[width=\hsize]{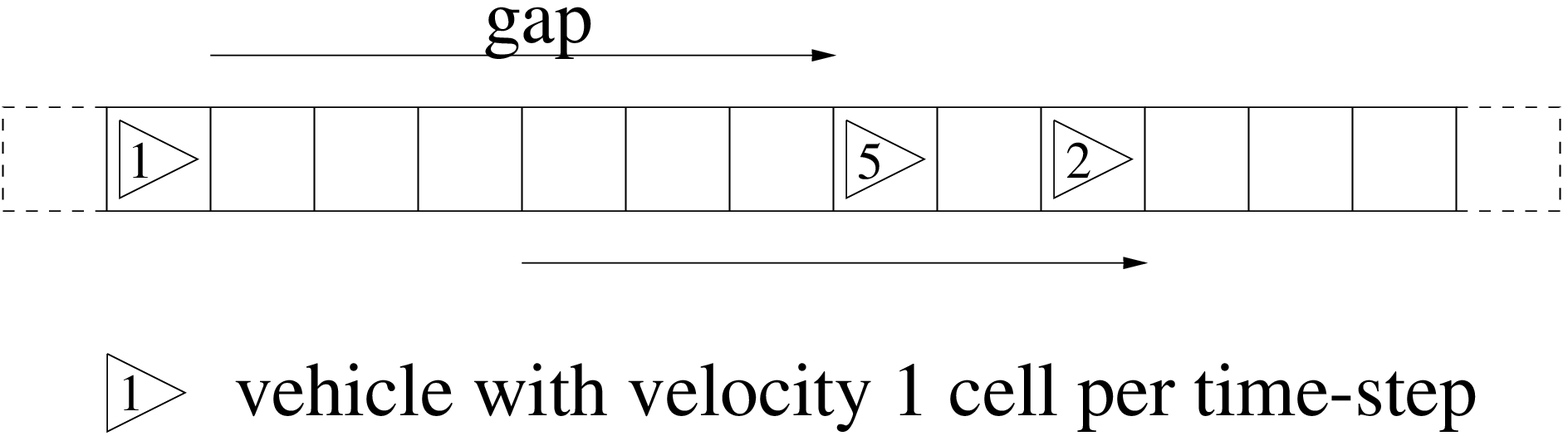}
\end{minipage}
\label{fig:gapdef}
}
}
\centerline{%
\subfigure[]{%
\begin{minipage}[c]{0.95\hsize}
\includegraphics[width=\hsize]{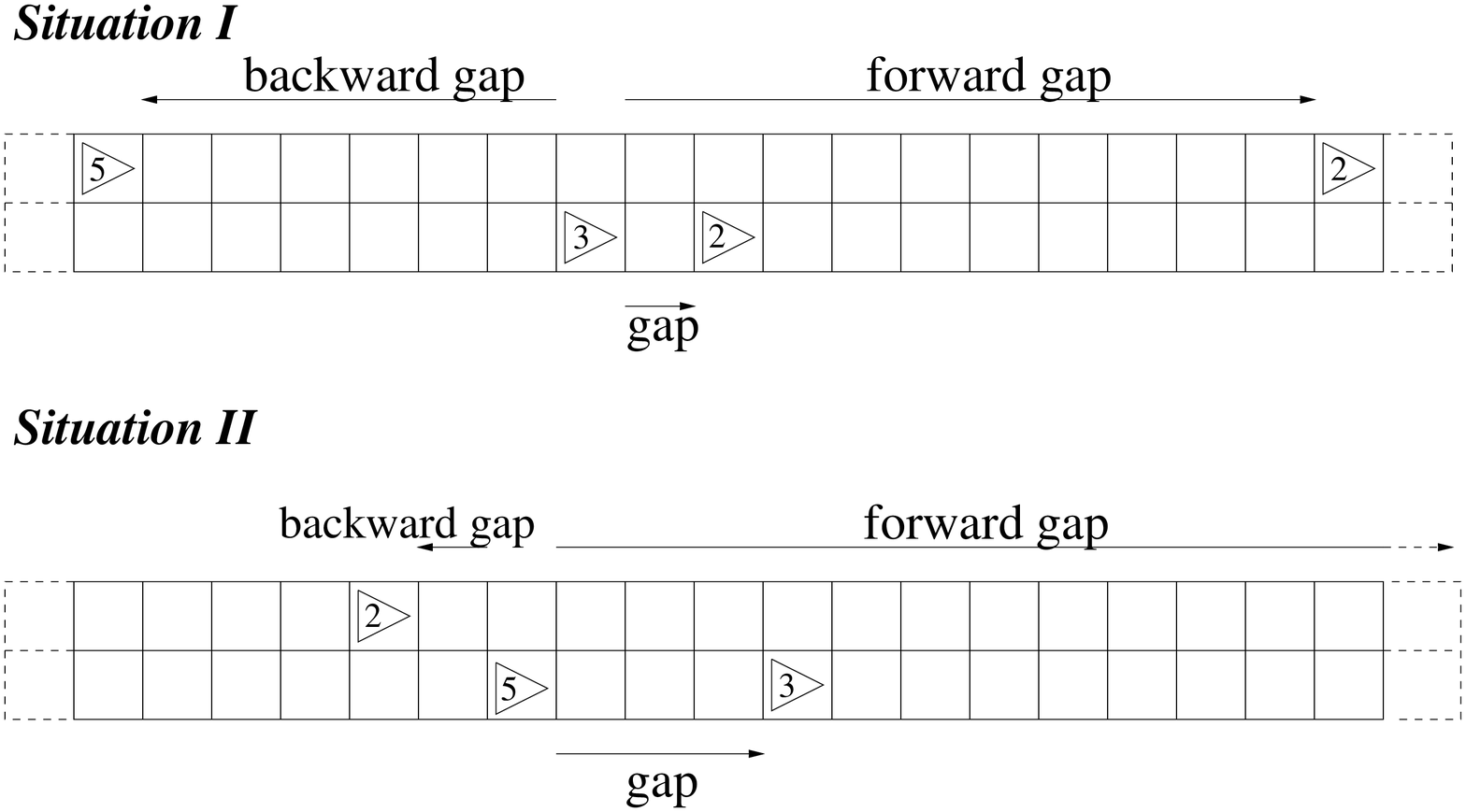}
\end{minipage}
\label{fig:lanechange}
}
}
\centerline{%
\subfigure[]{%
\begin{minipage}[c]{0.95\hsize}
\includegraphics[width=\hsize]{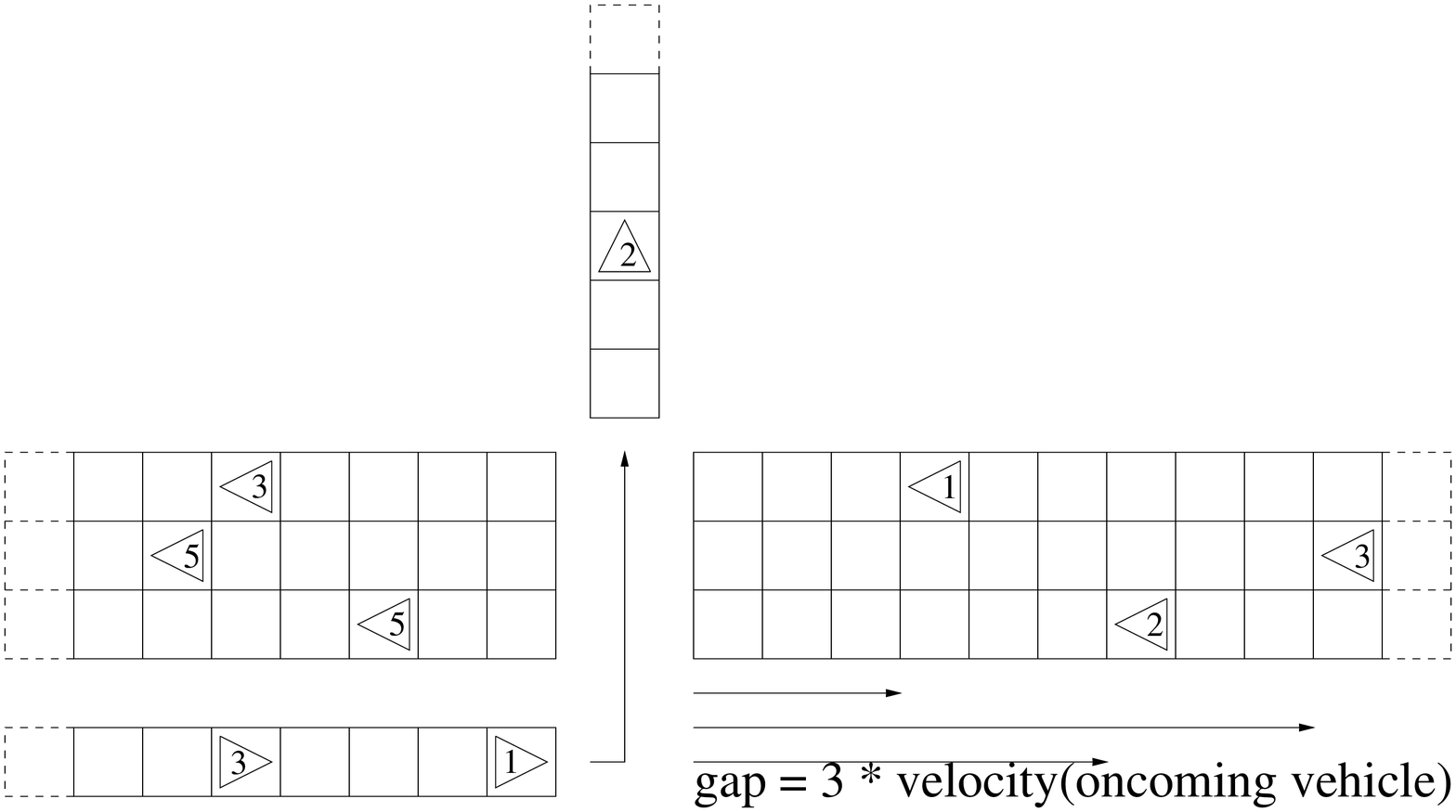}
\end{minipage}
\label{fig:leftturn}
}
}
\caption{%
Illustration of cellular automata driving rules (from
\cite{Nagel:Raney:w-schmid-spatial-planning}).
{(a)} Definition of $gap$ and one-lane update.
{(b)} Lane-changing.
{(c)} Left turn against oncoming traffic.
}
\label{fig:ca}
\end{figure}

\begin{figure}[t]
\centerline{%
\hfill
\begin{minipage}[c]{0.9\hsize}
\includegraphics[width=\hsize]{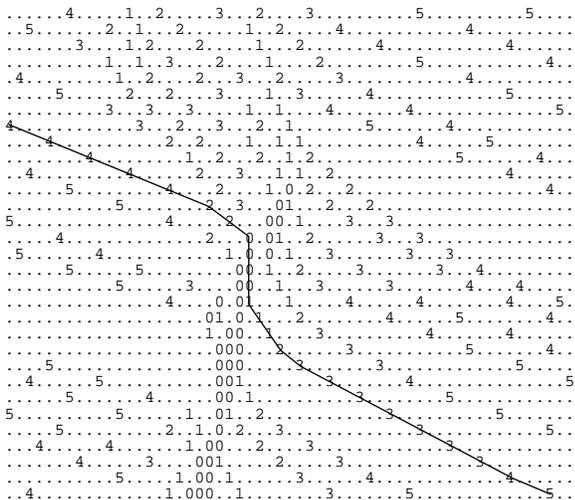}
\end{minipage}
\hfill
}
\caption{%
Emergence of a jam ``out of nowhere'' (``phantom traffic jam''),
simulated with a stochastic cellular automaton.
}
\label{fig:stca}
\end{figure}

Typical rules for lane changing (see Fig.~\ref{fig:lanechange})
include a reason to change lanes and a safety criterion for changing
lanes.
First, there needs to be a reason why
a vehicle wants to change lanes, for example that the other lane is
faster, or that it needs to get into a certain lane in order to make
an intended turn at the end of the link.  A possible rule to model the
first is ``check if the (forward) gap in the other lane is larger than
the gap in the current lane.''
Second, the vehicle needs to check that
there is really enough space in the destination lane.  A possible rule
is that the forward gap in the other lane needs to be larger than $v_t$, 
and the backward gap in the other lane needs to be larger than the
velocity of the oncoming vehicle.

Figure~\ref{fig:lanechange} illustrates the lane changing rules.  Only
lane changes to the left lane are considered.  In situation I, the
leftmost vehicle on the bottom lane will change to the left because
the forward gap on its own lane, 1, is smaller than its velocity, 3;
the forward gap in the other lane, 10, is larger than the gap on its
own lane, 1; the forward gap in the target lane is large enough; and
the backward gap is large enough.  In situation II, the second vehicle
from the right on the right lane will not perform a lane change
because the gap backwards on the target lane is not sufficient.

Due to the complexity of the dynamics, it is inconvenient to do both
lane changing and car following in one parallel update.  It is however
possible, and convenient for parallel computing, to do the update
in two completely parallel sub-timesteps: First, all lane changes are
executed in parallel based on information from time $t$, resulting in
intermediate information, at time $t\!+\!\frac{1}{2}$.  Then, all car
following rules are executed in parallel, based on information from
time $t\!+\!\frac{1}{2}$, which results in information for time
$t\!+\!1$.

\subsection{Intersections}
\label{sec:intersections}

The typical modeling substrate for transportation is the network,
which is composed of links and nodes that model roads and
intersections, respectively.
Since the full simulation of an intersection is rather complicated,
for large scale applications one attempts to reduce the complexity by
resolving all conflicts at the entering of the intersection.  Under
such a simplification, it is still possible to model things like left
turns against oncoming traffic, but the decision is made before the
vehicle enters the intersection, not inside the intersection.  In
consequence, complications inside the intersection, such as the
interesting dynamics of a large roundabout in Paris, are not modeled
by this approach.  Nevertheless, the simplified approach is quite
useful for large scale applications.

Given this, there are only two remaining cases for traffic: Protected
and unprotected turns.  A protected turn means that a signal schedule
takes care of possible conflicts, and thus the simulation does not
have to worry about them.  In this case, it is enough if the
simulation makes vehicles stop when a signal is red, and lets them go
when a signal is green.  

In unprotected turns, conflicts are resolved by the legal rules, not
by signals.  For example, a vehicle making a left turn needs to give
priority to all oncoming vehicles (Fig.~\ref{fig:leftturn}).  This
entails that one needs, for each turning direction, to encode which
other oncoming lanes have the priority.  Once this is done, a vehicle
that intends to do a certain movement only needs to check if there is
a vehicle approaching on any of these interfering lanes.  If so, the
vehicle under consideration has to stop, otherwise it can proceed.  As
said before, once the vehicle has entered the intersection, other
vehicles are no longer regarded.
Figure~\ref{fig:leftturn} shows an example of a left turn against oncoming traffic.  The turn
is accepted because on all three oncoming lanes, the gap is larger or
equal to three times the first oncoming vehicle's velocity.

\subsection{Modeling Queues}
\label{sec:queue-model}

Sometimes, even the relatively simple CA rules are computationally too
slow.  Another factor of ten in the execution speed can often be
gained by simplifying the dynamics even more.  In the so-called queue
model \citep{Gazis:queue,Gawron:queue}, links have no internal
structure any more, they can be considered as vehicle reservoirs.
These reservoirs have a storage constraint, $N_{max}$, which varies
from link to link, and is based on the physical attributes of the
link, such as length and number of lanes.  Once the storage constraint
is exhausted, no more vehicles can enter the link until some other
vehicles have left the link.

Vehicles can only leave the link if their destination link has space,
if they have spent the time it takes to traverse the link, and if the
flow capacity constraint is fulfilled.  The flow capacity constraint
is a maximum rate with which vehicles can leave the link.

Except for the storage constraint and its consequences, this is just a
regular M/M/1 queuing model.  The storage constraint does, however,
necessitate one adaptation of the model: Since destination links can
be full, there is now competition for space on the destination link.
A good option is to give that space randomly to incoming links, with a
bias toward incoming links with higher capacity
\citep{queue}.

With the queue model and using parallel computing, it is possible to
simulate a full day of all car traffic of all of Switzerland
(approx.~7~mio inhabitants) in less than 5~minutes \cite{queue}.
Although data movement is still a challenge, this makes it possible to
run completely microscopic (agent-based) studies of large metropolitan
areas \cite{ch}.

\section{Entropy Laws and Destination Choice}
\label{sec:entr-laws-dest}

The simulation of traffic in street networks requires
data on the number $V_{kl}$ of trips between origins $k$ and destinations $l$
as a function of the time of the day, the weekday etc. These origin-destination data
are relatively scarce and expensive. However, the number $Q_k = \sum_l V_{kl}$ of trips 
starting in a city $k$ or quarter of a city and the number $Z_l = \sum_k V_{kl}$ of trips 
ending in $l$ is relatively well-known, as this requires only $2n$ instead of $n^2$ data,
where $n$ denotes the number of dinstinguished origins/destinations. Apart from this,
$Q_k$ and $Z_l$ can be automatically measured by detectors, while the determination
of the entries $V_{kl}$ of the traffic flow matrix requires to obtain origin-destination pairs,
i.e. to ask drivers.
\par
Therefore, the relative frequencies $p_{kl} = V_{kl}/V$ from the origins $k$ to the
destinations $l$, where $V= \sum_{k,l} V_{kl}$,  are often determined via the ``most likely''
distribution under the constraints 
\begin{equation}
 \sum_l p_{kl} = Q_k/V = p_k \quad \mbox{and} \quad \sum_k p_{kl} = Z_l/V = p_l \, .
\label{like}
\end{equation}
This is obtained by minimizing the {\em ``information gain''}
\begin{equation}
 I = - \sum_{k,l} p_{kl} \ln \frac{p_{kl}}{b_{kl}} 
\label{take}
\end{equation}
compared to some ``natural distribution'' $b_{kl}$. 
This is analogous to maximizing entropy $H = - \sum_{kl} p_{kl} \ln p_{kl}$ in cases where 
$b_{kl}$ is an equidistribution. The only difference
is that (\ref{take}) takes into account a ``resistance function'' $b_{kl}$, which 
reflects the ``natural distribution'' as a function of the
distance and other behaviorally relevant variables in the absence of constraints like (\ref{like}). 
The appropriate specification of $b_{kl}$ will be discussed in Sec.~\ref{resist}.
\par
Instead of minimizing (\ref{take}) under the constraints (\ref{like}), it is easier to
minimize
\begin{eqnarray}
 L &=& - \sum_{k,l} p_{kl} \ln \frac{p_{kl}}{b_{kl}} \nonumber \\
 &+& \sum_k \lambda_k \left( p_k - \sum_l p_{kl} \right) \nonumber \\
 &+& \sum_l \mu_l \left( p_l - \sum_k p_{kl} \right) \, ,
\end{eqnarray}
where the additional terms vanish when the constraints (\ref{like}) are fulfilled.
Differentiation with respect to $p_{k'l'}$ gives
\begin{equation}
 - \ln \frac{p_{k'l'}}{b_{k'l'}} - 1 - \lambda_{k'} - \mu_{l'} = 0
\end{equation}
or
\begin{equation}
 p_{kl} = b_{kl} \underbrace{\mbox{e}^{-\lambda_k - 1/2}}_{=g_k} \, 
\underbrace{\mbox{e}^{-\mu_l - 1/2}}_{=h_l} \, .
\label{analog}
\end{equation}
The $2n$ Lagrange parameters $\lambda_k$ and $\mu_l$ are now determined in a way
that the $2n$ constraints (\ref{like}) are satisfied. This requires the use of
numerical iteration procedures. For an overview of numerical solution algorithms see 
Refs.~\cite{Lohse1,Lohse2,Lohse3}. These references also suggest ways to treat
the choice of destinations and transport modes simultaneously.

\subsection{Energy Laws and Resistance Function} \label{resist}

According to the last paragraphs, the distribution of traffic significantly depends on the
``natural distribution'' $b_{kl}$. 
Robert K\"olbl and one of us have investigated the electronically available empirical data of the UK
National Travel Survey during the years 1972--1998 \cite{eins}. When we distinguished
between different daily modes of transport $j$, we found that the average
modal travel time $\overline{t}_{j}$ per day and person remained almost
constant over the 27 years of observation,
despite variations by a factor 3.8 between different modes. More specifically,
the average travel times $\overline{t}_{j}$ were $40
$~min during a day with walking ($j=\mbox{w}$), without any usage
of other means of transport, $42%
$~min for cyclists ($j=\mbox{c}$), $67%
$~min for stage bus users ($j=\mbox{b}$), $75%
$~min for car drivers ($j=\mbox{d}$), $59%
$~min for car passengers ($j=\mbox{p}$), and $153%
$~min for train passengers ($j=\mbox{t}$).
\par
\begin{figure}[hptb]
\begin{center}
\hspace*{-2mm}\includegraphics[width=6.82cm, angle=-90]{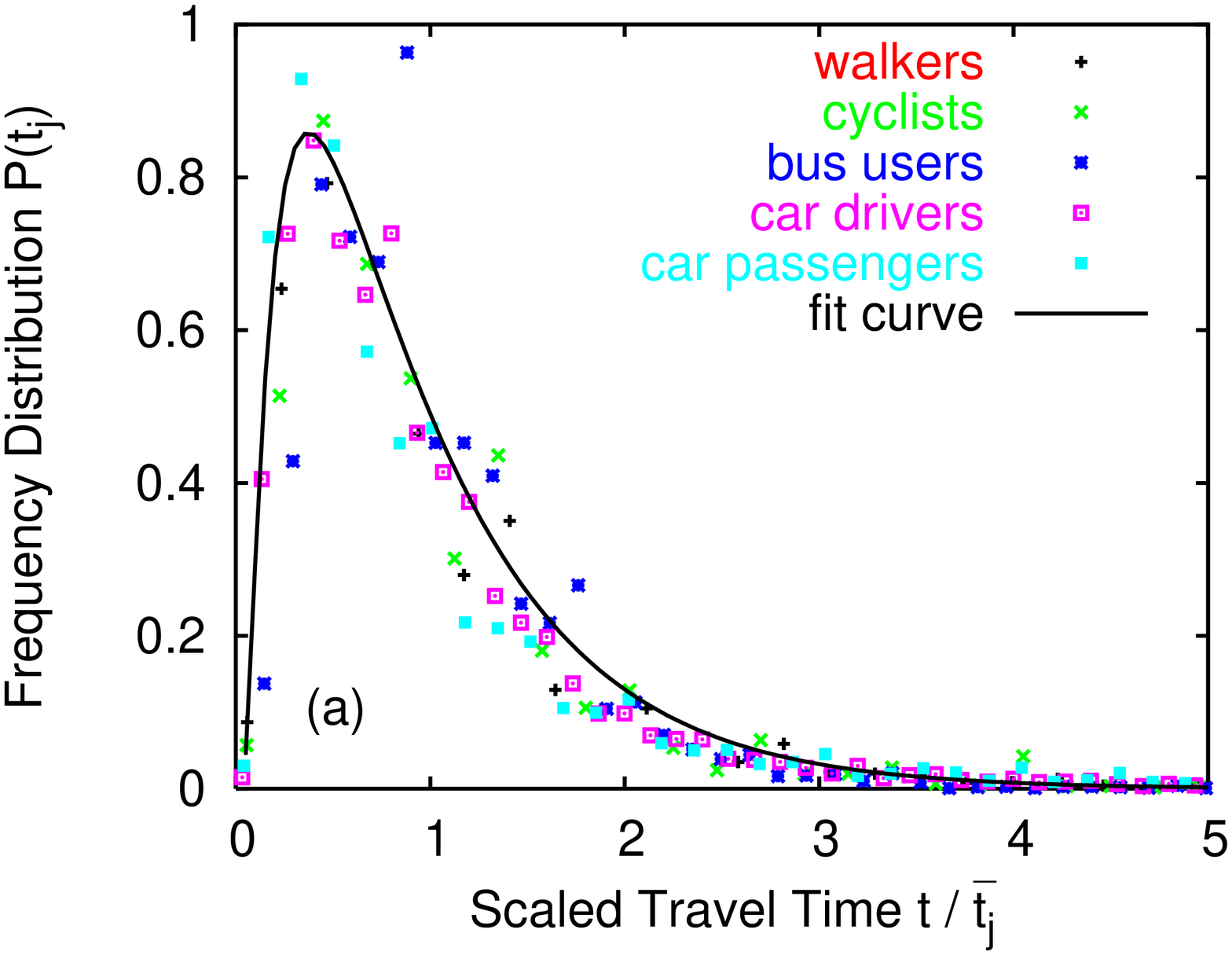}\\
\hspace*{-3mm}\includegraphics[width=6.6cm, angle=-90]{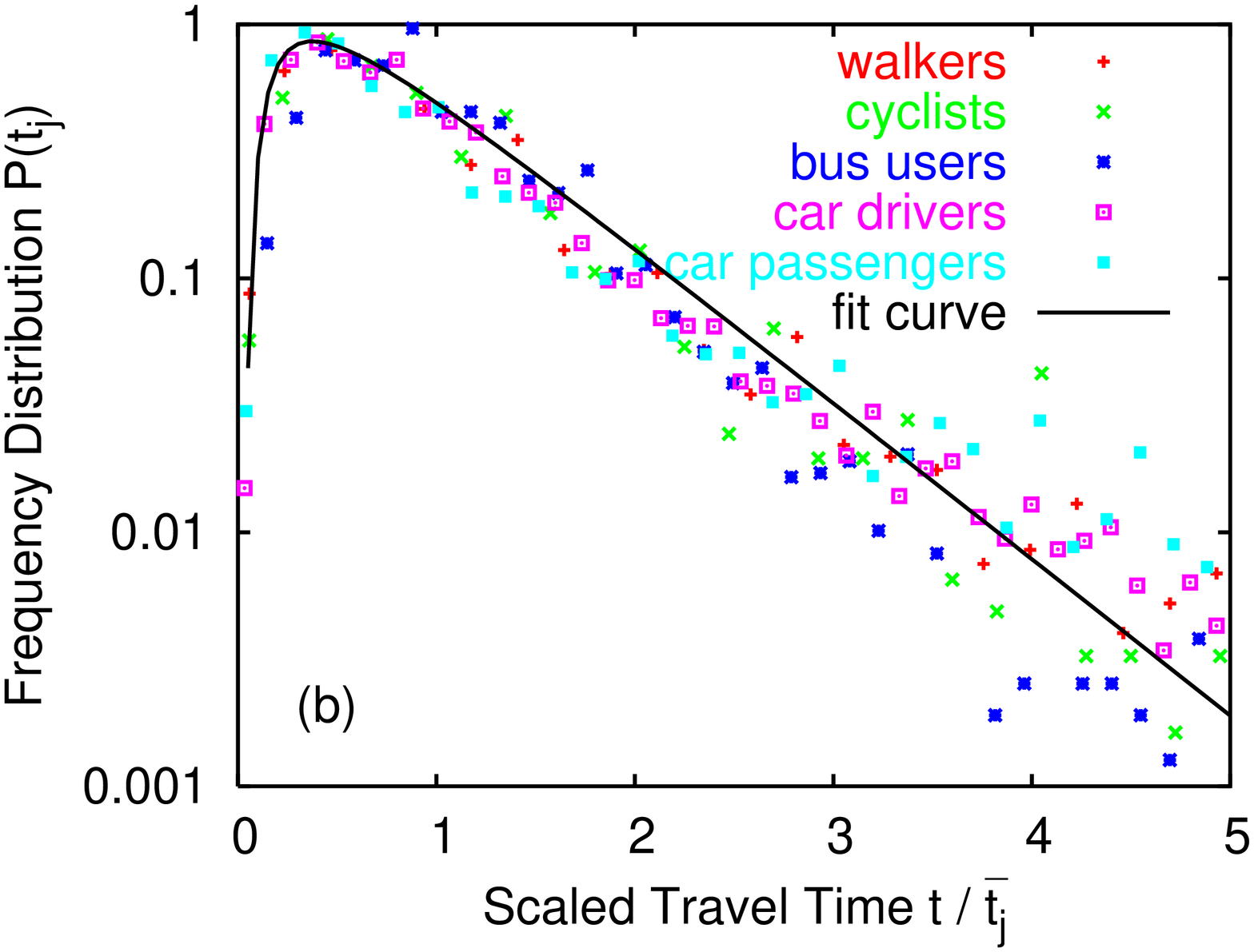}
\end{center}
\caption[]{Scaled time-averaged travel-time distributions for
different modes of transport (a) in linear and (b) in semilogarithmic representation.
Within the statistical variation (and rounding errors for small frequencies,
which are magnified in the semilogarithmic plot),
the mode-specific data could all be fitted by one universal curve (\ref{pt})
with $\alpha= 0.2$, $\gamma= 0.7$, and normalization constant
$N(\alpha,\gamma) = 2.5$. The few railway data were not significant 
because of their large scattering. (From \cite{robert}.) \label{travel}}
\end{figure}

When we scaled the $t$-axis of the modal travel-time distributions
$P_{j}(t)\,dt$ by the average travel times $\overline{t}_{j}$, we discovered
that, within the statistical variation, the resulting distributions collapsed onto
one single curve
\begin{equation}
P(t_{j})\,dt_{j}\approx N\exp\left(-\frac{\alpha}{t_{j}} - \frac{t_{j}}{\gamma} \right) \, dt_{j}
\label{pt}
\end{equation}
with $t_{j}=t/\overline{t}_{j}$, two fit parameters $\alpha$ and $\gamma$, and
the normalisation constant $N=N(\alpha,\gamma)$ (see Fig.~\ref{travel}a). This implies a
universal law of human travel behaviour and the resistance function or natural distribution
\begin{equation}
 b_{kl} = N\exp\left( - \frac{\alpha}{w_{kl}} - \frac{w_{kl}}{\gamma} \right) \, ,
\label{be}
\end{equation}
where $w_{kl}=t_{kl}/\overline{t}_{j}$ is the travel time $t_{kl}$ from $k$ to $l$ divided by
the average travel time $\overline{t}_j$ for transport mode $j$.
\par
In the semi-logarithmic representation $\ln P(t_{j})=\ln N-\alpha/t_{j}-t_{j}/\gamma$ of (\ref{pt}), the term
$\alpha/t_{k}$ is relevant only for short time scales up to $t_{j}\approx 0.5$,
while the linear relationship $\ln P(t_{j})=\ln N-t_{j}/\gamma$ dominates
clearly over a wide range of scaled travel times $t_{j}$ (see Fig.~\ref{travel}b).
Therefore, we suggest the following interpretation: The dominating term $P_{j}
(t_{j})\,dt_{j}\propto\exp(-t_{j}/\gamma)\,dt_{j}$ corresponds to the
canonical energy distribution
\begin{equation}
P^{\prime}(E_{j})\,dE_{j}=N^{\prime}\exp(-E_{j}/\bar{E})\,dE_{j} \, ,
\label{behavior}
\end{equation}
where $E_{j}= p_j t$ with $p_j = \bar{E}/(\gamma\overline
{t}_{j})$ stands for the energy consumed by the physical activity of travelling
with transport mode $i$ for a time period $t$.
The term $P_{j}(t_{j})\propto\exp(-\alpha/t_{j})$ possibly represents the additional
amount of energy required for the mobilization of
short trips, i.e. the Simonson effect \cite{drei}. 
With an average travel energy budget of $\bar{E}=615$~kJ, this
interpretation is consistent with values of the energy consumption 
$p_j$ per unit time for certain activities obtained by ergonomic measurements of 
the related O$_{2}$-consumption \cite{vier1,vier2,vier3}. 
\par
Based on entropy maximization, the canonical distribution can be
shown to be the most likely distribution, given that the average energy
consumption $\bar{E}$ per day by an ensemble of travellers
is fixed for the area of investigation. This agrees well with the investigated
data and generalizes the concept of a constant travel-time budget
\cite{f1,f2,f3,f4,f5,f6,f7,f8}. 
In addition, the above theory can be combined with trip distribution
models \cite{MNL,s1,s2,s3,s4,s5}, e.g.\ the multinomial logit model
\begin{equation}
P_{j}^{\ast} = \frac{\mbox{e}^{\beta' V_j}}{\sum_{j'} \mbox{e}^{\beta' V_{j'}}} \, ,
\label{eq:MNL}
\end{equation}
where $\beta'$ is a parameter (see Sec.~\ref{sec:MNL}) and the mode-dependent energy consumption enters the utility
function $V_j$ as a (negative) cost term, which was shown to be a
significant variable of human travel behaviour.
\par
In summary, when daily travel time distributions by different modes of transport
such as walking, cycling, bus or car travel are appropriately scaled, they
turn out to have a universal functional relationship. A closer investigation reveals
a law of constant average energy consumption by the physical activity of
daily travelling. The advantage of the behavioral law (\ref{pt}) is its expected long-term validity
over more than 25 years even under changing conditions. It will, therefore, prove to be important for
urban, transport, and production planning: Compared to previous models,
it facilitates improved quantitative conclusions about trip distributions, 
modal splits and induced traffic after the more reliable determination of fewer parameters, 
which are constant over typical planning horizons. 
It also helps to assess the increase in the  acceptability of public transport, when the
comfort of travel is improved,  to forecast the change in travel behavior in an aging society,
to predict the usage of new modes of transport,
and to  estimate potential market penetrations of new travel-related
products.

\section{Activity Patterns}
\label{sec:activity-patterns}

Sometimes, the models outlined in the previous sections are not
sufficient to model demand generation for traffic.  The first obvious
shortfall is that they do not say anything about a time-of-day
dynamics.  This is sometimes introduced by assuming that all trips
from residential to commercial locations represent morning traffic,
and the afternoon/evening peak is then obtained by transposing the
trip matrix.

Often, even that is not enough.  This happens typically when there are
significant correlations between travelers' attributes and their
decisions.  For example, the number of trips starting in a zone in
general does not just depend on the number of residents, but also on
their income or age; destination choice in general depends on vehicle
availablity; the acceptance of a new mode of transport in general
depends on the question if people can reorganize their entire day
plans so that the new facility can be efficiently integrated.

\note{make agent based}

A possible way out of this is to make the entire approach completely
microscopic, or agent-based.  This means that throughout the whole
simulation, all travelers are maintained as individuals, and
individual behavioral models are applied at all steps of the
decision-making.  For example, synthetic ($=$ simulated) travelers may
plan their activity patterns (e.g.\ home -- work -- shop -- home --
leisure -- home), then the corresponding locations, then the modes of
transportation, and finally the precise timing of their activities.
The advantage of this approach over the ones discussed so far is that,
at each step of the decision-making, all ``internal'' information of
the agent is available. 

\note{genealogy}

Such an approach draws from many areas of science.  Clearly,
psychology is involved because it describes the behavior of human
individuals.  Yet, since the goal is not to describe each individual
traveler correctly but to get correct aggregate results, economics and
the social sciences are involved as well, and there is in fact a
community of ``travel behavior research'' \citep{iatbr00-book} which
applies these principles to traffic. However, the approach to go to
microscopic principles when an analysis on the macroscopic level fails
is also deeply rooted in physics, and in fact, both the computing
methods to large-scale problems and the general intuition that
simple (and therefore individually wrong) microscopic models may lead
to correct macroscopic results come to a large extent from physics.

\note{preview}

Given its highly interdiscplinary nature plus the fact that only
recently data and computing methods became available to do
activity-based demand generation for large urban areas, it is
understandable that the area is in rapid flux.  There is one method,
discrete choice theory \citep{Ben-Akiva:book}, that is arguably better
established than others, and it will be described in a bit more detail
in the following.  However, there are some drawbacks to that theory,
and it is questionable whether it will be possible to correct them
within the framework of that theory.  For that reason, alternative,
less well established approaches will also be discussed.

\subsection{Discrete Choice Theory}

Assume, for simplicity, a situation where an agent has a choice between
several alternatives, $j$, for example between several locations for
the same activity.  Also assume that there are utilities $U_j$
associated with those alternatives.  Discrete choice theory starts
from two assumptions:
\begin{itemize}

\item It is possible for the observer/modeler to measure/predict some
part of the utilities.  These measurable parts are called ``systematic
utilities'' $V_j$.

\item The differences between the $V_j$ and the $U_j$ are random and
statistically independent.  These differences are denoted as $\eps_j$.

\end{itemize}
With these assumptions, one obtains essentially the following
algorithm:
\begin{enumerate}

\item Compute the $V_j$ based on your observations.  For example, one
might have observed that one location is accessible only via some
steep stairs, and therefore avoided by physically less fit persons.

\item Add the random components: $U_j = V_j + \eps_j$.

\item Select the alternative that \emph{now} has the highest utility.
Note that, because of the random component, it can well be that the
alternative with a smaller $V_j$ in the end ``wins''.

\end{enumerate}
There are at least two different interpretations where the random
parts $\eps_j$ come from: Economic theory states that economic agents \emph{always}
select the option with the largest utility, and the difference between
the observed/predicted utilities $V_j$ comes from so-called unobserved
attributes of the agent which are thus unknown to the
observer/analysis.  Psychology, in contrast, might say that
there is just a random component to people's behavior.

\subsection{Multinomial Logit Model (MNL)} \label{sec:MNL}

The question now is how to obtain real-world numbers for the $V_j$ and
the $\eps_j$. The typical way to progress is in two steps: (i)~make
some assumptions about the $\eps_j$; (ii)~assume that the $V_j$ are
additive in their contributions and do a maximum likelihood estimation
of the parameters.

\paragraph{The random components.} 
A conceptually easy way to proceed is to assume that the $\eps_j$ are
normally distributed.  The result of this is called a
\emph{(multinomial) probit model}.  Its solution, however, cannot be
written down in terms of simple functions.  Another path is to assume
that the $\eps_j$ follow so-called Gumbel-distributions with zero
mean, i.e.\ their generating function is $F(\epsilon_j) = \exp[ - \exp(
- \mu_j \, \epsilon_j ) ]$, where $\mu_j$ is a width parameter.  In
this case, the result is that the probability for agent $i$ to select
option $j$ follows the Boltzmann distribution
\begin{equation}
p_{i,j} = \frac{e^{\beta' \, V_{i,j}}}{\sum_k e^{\beta' \,  V_{i,k}}} \ .
\label{eq:mnl-2}
\end{equation}
This is known as the \emph{(multinomial) logit model}, cf. 
Eq.~(\ref{eq:MNL}).  $\beta'$ results from the ``widths'' $\mu_j$ of the
distributions.  It models how ``rational'' the agents behave with
respect to what is observed: a small $\beta'$ makes the agents nearly
random with respect to $V_j$; for $\beta' \to \infty$ one obtains an
agent that always takes the option $j$ with the best $V_j$.

\paragraph{The systematic components.}  
In order to make further progress, one has to make an assumption about
the functional form of the $V_j$, and its free parameters.  The most
typical assumption is that the $V_j$ are linear in their components.
For example, the utility for agent $i$ to go swimming at location $j$
might be
\begin{equation}
V_{i,j} = \beta_T \, T_{i,j} + \beta_C \, C_{i,j} + \beta_G \, G_i \,
\label{eq:attributes}
\end{equation}
where $T_{i,j}$ is the travel time to get there, $C_{i,j}$ is the
monetary cost, and $G_i$ is $i$'s gender (e.g.\ 0 for male, and 1 for
female).  Normally, $\beta_T$ and $\beta_C$ will be negative.  A
positive $\beta_G$ would express that, all other things being equal,
females like to go swimming more than males.  Note that attributes of
the alternative as well as attributes of the person can occur, and
that attributes can be continuous or discrete.

One now proceeds by asking a large number of people about their
individual $T_{i,j}$, $C_{i,j}$, $G_i$, plus the decision that they
actually made (revealed preference) or would make (stated preference)
in the given situation.  Maximum likelihood estimation then finds
coefficients $\beta_X$ such that the probability to obtain the
decisions from the survey is maximized ($X$ stands for the
different possible indices).  More precisely, it finds
products $\beta' \, \beta_X$, meaning that the $\beta'$ from
Eq.~(\ref{eq:mnl-2}) is already included.
Note that it is \emph{not} necessary for Eq.~(\ref{eq:attributes}) to
be linear; something like $\beta_{TT} \, T_{i,j}^2$ or $\beta_{CI} \,
C_{i,j}/I_i$ is possible (where $I_i$ is $i$'s income).
In the past, the utility function $V_{i,j}$ had to be linear in the
$\beta_X$ (because of the need for a partial derivative for the
maximum likelihood estimation), but that requirement is now lifted
\cite{biogeme}.

\subsection{Extensions}

Multinomial logit models (MNLs) have a property called
\emph{independence from irrelevant alternatives} (IIA).  It means
that taking an alternative out of the choice set does not change the
relative probabilities: $p_{i,j}/p_{i,k}$ does not depend on a third
alternative $m$.  This has sometimes odd consequences, typically
explained in terms of the so-called ``red bus blue bus'' paradoxon.
Assume that a traveler has the option to take a car (option~1) or to
take the bus, and that she takes the bus with 50\% probability.  Now
assume that the bus route is served by two buses, one of them red
(option~2) and one of them blue (option~3), and both of them depart at
exactly the same time, they have exactly the same service quality, and
they are always nearly empty.  Estimation of an MNL would result in a
25\% probability for the red bus and a 25\% probability for the blue
bus.

So far, this is all fine.  However, now assume that the blue bus gets
taken out of service.  The model for our traveler now predicts that
she uses the car with probability $2/3$, and the red bus with
probability $1/3$.  This is due to the fact that the ratio of the
probabilities of option~1 and option~2 has to remain unchanged.

The problem sits in the fact that for this example, the random
components $\eps_2$ and $\eps_3$ are not independent.  One way to deal
with this problem is to make the model hierarchical (``nested''),
where first the choice is between car and bus, and then about the type
of bus.  This reduces computational complexity, but demands that the
analyst specifies the hierarchical structure.  Another option is to
use models in which correlations between the random variables are
included as covariances.  The results are more sophisticated than for
the MNL model, but with increasingly better computers they enter the
domain of being useful for practical cases. 

Another extension concerns the parameters $\beta_X$, where $X$ again
stands for the different possible indices.  For example, in
Eq.~(\ref{eq:attributes}), it would make sense to assume that the
ratio $\beta_T/\beta_C$, related to the ``value of time'', depended on
income.  The two options in this situation are (i)~to assume an
explicit income-dependent model such as $V_{ij} = \beta_{T} T_{i,j} +
\beta_{CI} \, C_{i,j}/I_i ...$ (where once more $\beta_T$ and
$\beta_{CI}$ would be expected to be negative) or (ii)~to
assume random taste variations across the population, which means that
the coefficients $\beta_X$ become random variables, with accordingly
more free parameters to be estimated.

Software for the estimation of the models described in this section
and pointers to relevant literature can be found at
\cite{biogeme}.

\subsection{Discrete Choice Theory for Activities}

Existing discrete choice models for the generation of daily activities
use the nested MNL approach, which translates into a hierarchical
decision tree: First a traveler decides on the daily pattern, then on
activity locations, then on modes of transportation, and then on exact
times \cite{Bowman:thesis,Bowman:etc:Portland:acts}.  A problem is
that decisions on the higher level depend on decisions on the lower
levels; for example, the decision about the inclusion of an activity
depends on how close its location is to the locations of other
possible activities.  The practical approach to that problem is to
first compute the lower level options and then pass them on as
``expected'' utilities to the higher levels.  For example, the utility
of doing a trip both by car and by bus is computed, and (roughly) the
maximum of both is passed on to the location choice module.

\subsection{Alternative Methods for the Generation of Activity Patterns}

In spite of the random term, discrete choice theory assumes extremely
rational agents.  In fact, a traveler modeled by a discrete choice
model for activity planning will consider the utilities of \emph{all}
different options before making a decision.  A consequence of this is
that the computation is in fact \emph{much more expensive than finding
the best option}, since for a best option, suboptimal branches in the
search tree can be pruned, while for discrete choice theory, all
branches need to be followed to the leaves.
In addition, discrete choice models for human activity planning need
of the order of several hundred free parameters, and it is
questionable how well so many parameters can be estimated from
surveys. 

For those reasons, it makes sense to look at alternative models.  For
example, one can make models where travelers do not look forward on
the time axis \cite{Kitamura:CO2}.  Such a model is much easier to
calibrate, but it is not sensitive to changes in the time structure:
For example, the model will not make people get up later in the
morning if opening times of shops are extended in the evening.  An
improvement of this is the use of Q-learning, where agents learn, by
doing the same day over and over again, to backpropagate the temporal
effects to the decision points \cite{q4acts}.  Again, other models are
entirely based on rules \cite{ALBATROSS} or on genetic algorithms
\cite{ga4acts}.  A newer development is the use of mental maps, where
the agent remembers which parts of the system it knows
\cite{Unger:phd,arentze:mental-map:iatbr03,dkistler:masters}.

\section{Learning, Feedback, and Evolutionary Game Theory}
\label{sec:learning-feedback}

Neither entropy-based destination choice
(Sec.~\ref{sec:entr-laws-dest}) nor complete activity patterns
(Sec.~\ref{sec:activity-patterns}) nor route choice
(Sec.~\ref{sec:route-choice}) can be sensibly modeled without
including the effect of learning or adaptation.  The simple reason for
this is that there are nonlinear circular dependencies: Human plans
depend on (expected) congestion, but congestion is the result of (the
execution of) plans.

The obvious solution for this is any kind of relaxation method, i.e.\
some version of ``make initial plans -- compute congestion -- adjust the
plans -- re-compute congestion'' etc.  Another interpretation of the
same approach is that it models human learning from one day to the
next.

It is interesting to note that there is actually some theory available
to describe this process.  For example, if all agents always play
``best reply'' (i.e.\ the best answer to the last iteration), and if
the iterative dynamics converges to a fixed point attractor, than that
fixed point is also a {\em Nash Equilibrium} (NE) \cite{Hofb:Sigm:book}.
For the traditional method of transportation planning, called static
assignment (similar to flows in electric networks except that
particles know their destinations), one can actually show that under
some assumptions the NE is unique (in the link flows), and therefore
the iterative process does not have to be explicitly modeled
\cite{Sheffi:book}.  For the more complicated models discussed in this
paper, no such mathematical statement is available, and therefore
multiple fixed point attractors could be possible, or the iterative
process could converge to a periodic or chaotic attractor.  If the
models are stochastic, then under mild assumptions they converge to a
fixed steady-state state-space density
\cite{Cascetta:Cantarella:day2day}.  In practice, only a small number
of iterations is feasible, and effects such as broken ergodicity
\cite{Palmer:broken:ergodicity} need to be taken into account.  A very
illustrative example of these different types of possible learning
dynamics will be provided in Sec.~\ref{sec:route-choice}.

Real world scenarios are even more complicated.  Decisions of
different agents may happen on different time scales, which depending
on their hierarchical order leads to different solutions
\cite{Zuylen:Taale:tgf03,Nagel:etc:sqrt-utl}.  This is related to the
issues of sequential games in game theory.  Agents may have limited
information, or they may not even use ``best reply'' at all, but
rather some {\em constraint satisficing method,} in which they improve
until they are satisfied.  Technically, this means that some important
real world results may lie in the transients rather than in the steady
state behavior.  A research agenda for the future needs to clarify
these questions, and needs to identify the limits of predictability
for such systems.

\section{Route Choice}
\label{sec:route-choice}

Once the activity pattern of an individual is selected (including the destination $j$ and the mode of 
transport $k$ \cite{Lohse1,Lohse2,Lohse3}),
one of the alternative routes $m$ to the destination needs to be chosen. This is typically done by maximizing the
utility $U_k$, which in most cases comes down to minimizing the overall travel time. 
As a consequence, the fastest connection is filled with vehicles, until the traffic density has increased
so much that their travel times become comparable with the travel time on an alternative road with greater length
or smaller speed limit. In this way, a distribution of traffic is produced, in which the travel times of all
alternative routes are the same. This distribution is called the {\em Wardrop equilibrium} \cite{Wardrop}.
However, does this distribution really describe the route choice behavior of drivers?
The following paragraphs will show that the Wardrop equilibrium systematically underestimates
travel times, as drivers do not match 
the ideal traffic distribution due to coordination problems.
\begin{figure}[htbp]
\begin{center}
\includegraphics[width=8.5cm, angle=0]{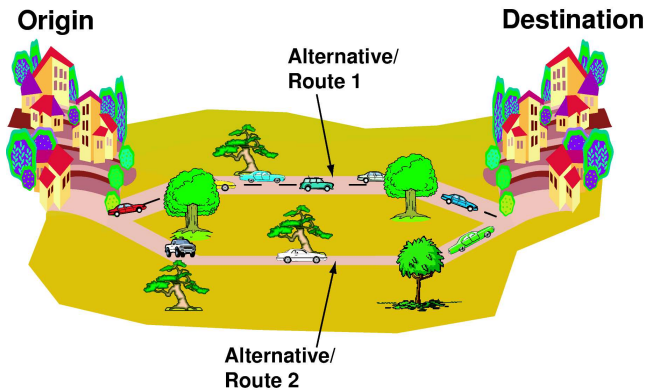}
\end{center}
\caption[]{Schematic illustration of the day-to-day route choice scenario
(from \cite{Control,PhysJ,Bonn}). Each day,
the drivers have to decide between two alternative routes, 1 and 2. Note that, due to the
different number of lanes, route 1 has a higher capacity than route 2. The latter is, therefore,
used by less cars.\label{illu}}
\end{figure}

\subsection{Decision Experiments}

\begin{figure}[htbp]
\begin{center}
\includegraphics[width=8cm, angle=90]{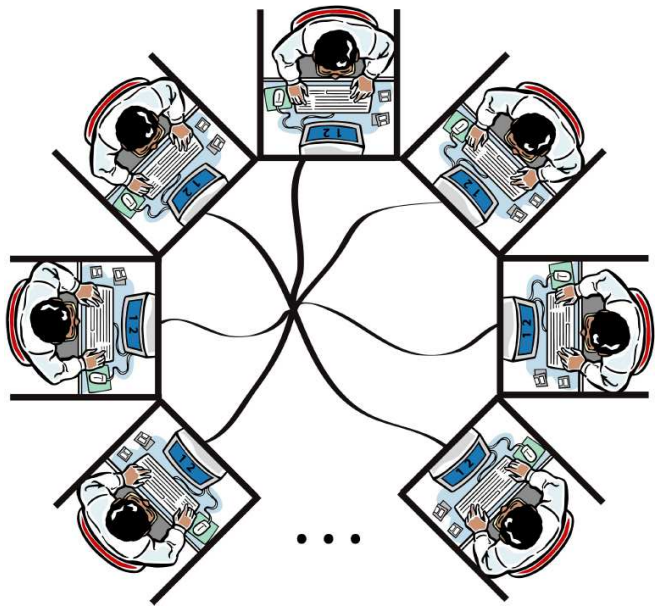}
\vspace*{-4mm}
\end{center}
\caption[]{Schematic illustration of the decision experiment
(from \cite{Control,Bonn}). Several test persons
take decisions based on the aggregate information their computer displays. The 
computers are connected and can, therefore, exchange information. However,
a direct communication among players is suppressed.
\label{compu}}
\end{figure}

\begin{figure}[htbp]
\begin{center}
\begin{picture}(13.5,17)(0,0.4) 
\put(-0.35,17.1){\includegraphics[height=13.8\unitlength, angle=-90]{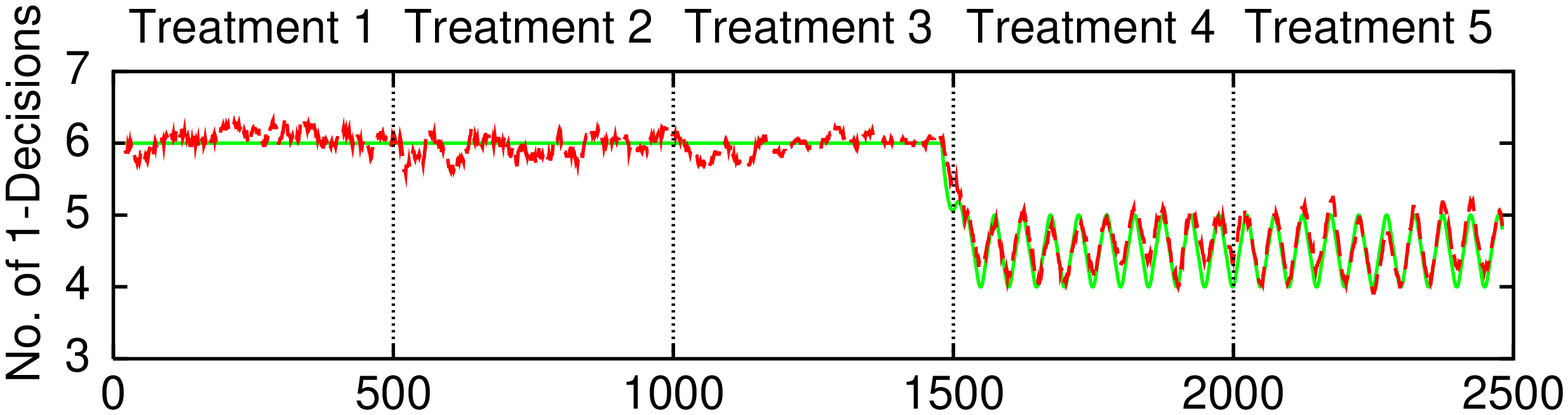}}
\put(-0.35,13.7){\includegraphics[height=13.8\unitlength, angle=-90]{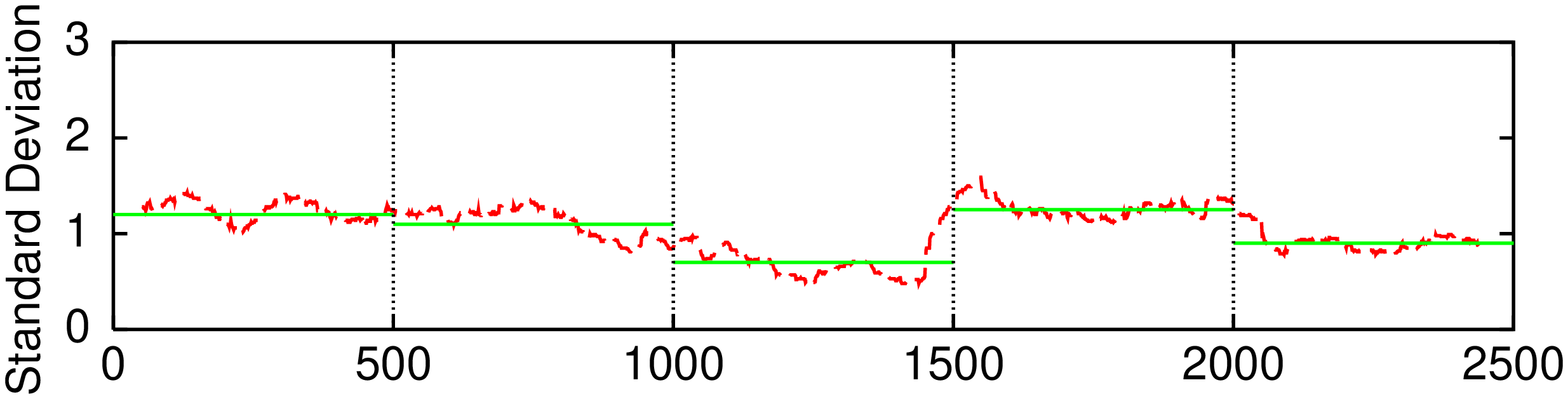}}
\put(-0.35,10.4){\includegraphics[height=13.8\unitlength, angle=-90]{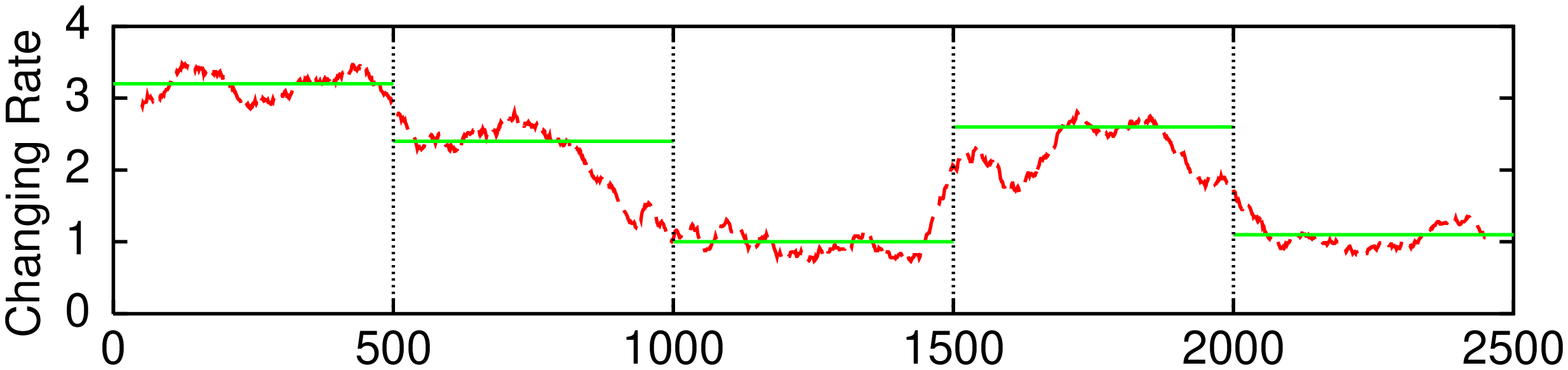}}
\put(-0.35,7.1){\includegraphics[height=13.8\unitlength, angle=-90]{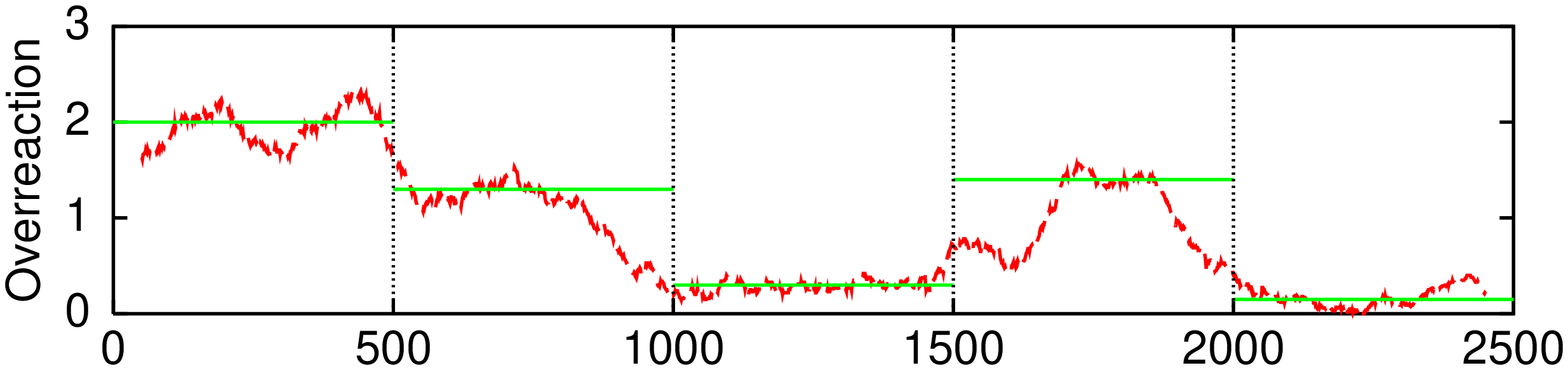}}
\put(-0.35,3.8){\includegraphics[height=13.8\unitlength, angle=-90]{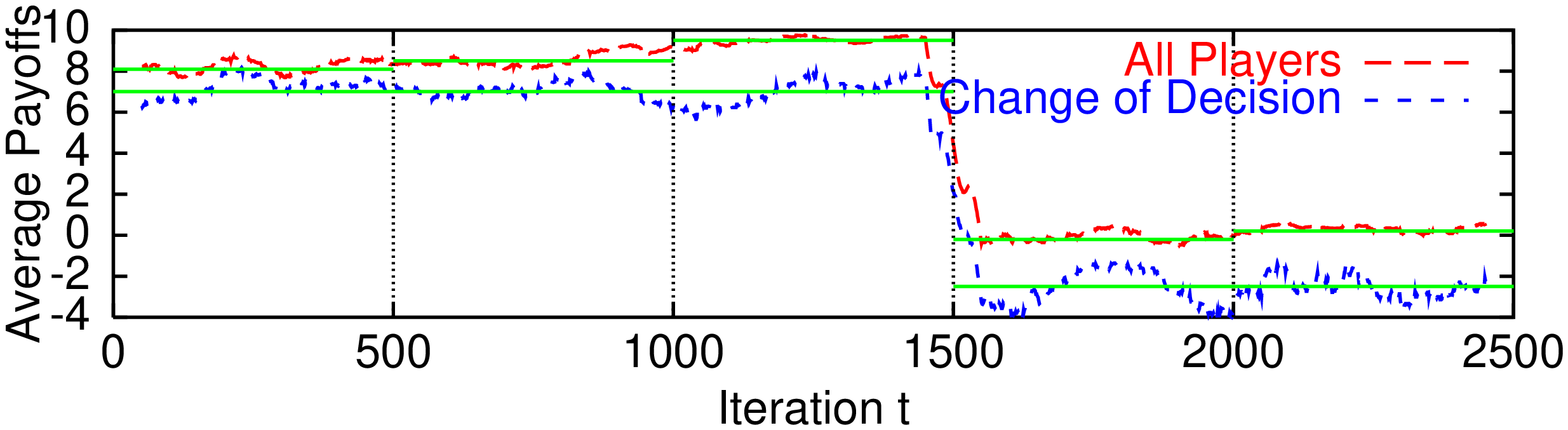}}
\put(1.1,14.25){\footnotesize\sf{(a)}}
\put(1.1,11){\footnotesize\sf{(b)}} 
\put(1.1,7.7){\footnotesize\sf{(c)}} 
\put(1.1,4.4){\footnotesize\sf{(d)}} 
\put(1.1,1.1){\footnotesize\sf{(e)}} 
\end{picture}
\end{center}
\caption[]{Overview of treatments 1 to 5 (with $N=9$ and payoff parameters $P_2^0=28$,
$P_1^1 = 4$, $P_2^1 = 6$, and $P_1^0 = 34$ for $0 \le t \le 1500$, but a zig-zag-like
variation between $P_1^0 = 44$ and $P_1^0 = -6$ with a period of $50$ for $1501 \le t \le 2500$):
{(a)} Average number of decisions for
alternative 1 (fluctuating curve) compared to the user equilibrium (smooth curve), 
{(b)} standard deviation of the number of 1-decisions from the user equilibrium,
{(c)} number of decision changes from one iteration to the next one,
{(d)} overreaction, i.e., difference between the actual number of decision changes 
(changing rate) and the required one (standard deviation),
{(e)} average payoff per iteration
for players who have changed their decision and for all players. The latter increased
with a reduction in the changing rate, but normally stayed below the 
payoff in the user equilibrium (which is 1 on average in treatments 4 and 5, otherwise 10). 
This payoff loss is caused by the overreaction in the system. 
The displayed moving time-averages
[(a) over 40 iterations, (b-e) over 100 iterations] illustrate the systematic
response to changes in the treatment every 500 iterations. Smooth lines in (b)--(e) 
are estimates of the stationary values after the transient period, while time periods around 
the dotted lines are not significant. Compared to treatment 1,
treatment 3 managed to reduce the changing rate 
and to increase the average payoffs 
(even more than treatment 2 did). 
These changes were systematic for {\em all} players. 
In treatment 4, the changing rate and the standard deviation went up,
since the user equilibrium changed in time. The user-specific recommendations in treatment 5 
could almost fully compensate for this and managed to reach the minimum overreaction 
in the system. The above conclusions
are also supported by additional experiments with single treatments. (After \cite{Control,Bonn}.)}
\label{Fig7}
\end{figure}

\begin{figure}
\begin{center}
\begin{picture}(13.5,11.5)(0,7)
\put(-0.25,18.5){\includegraphics[width=6.05\unitlength, angle=-90]{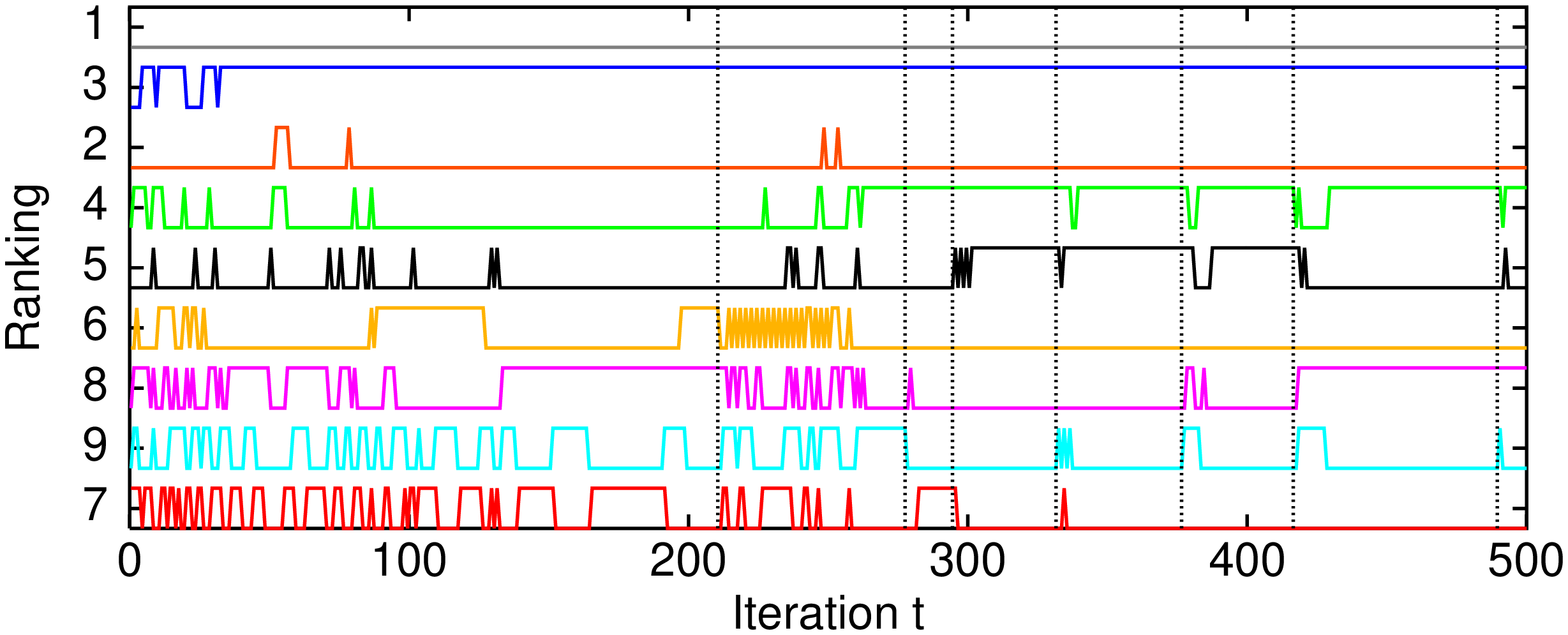}}
\put(-0.25,12.3){\includegraphics[width=6.05\unitlength, angle=-90]{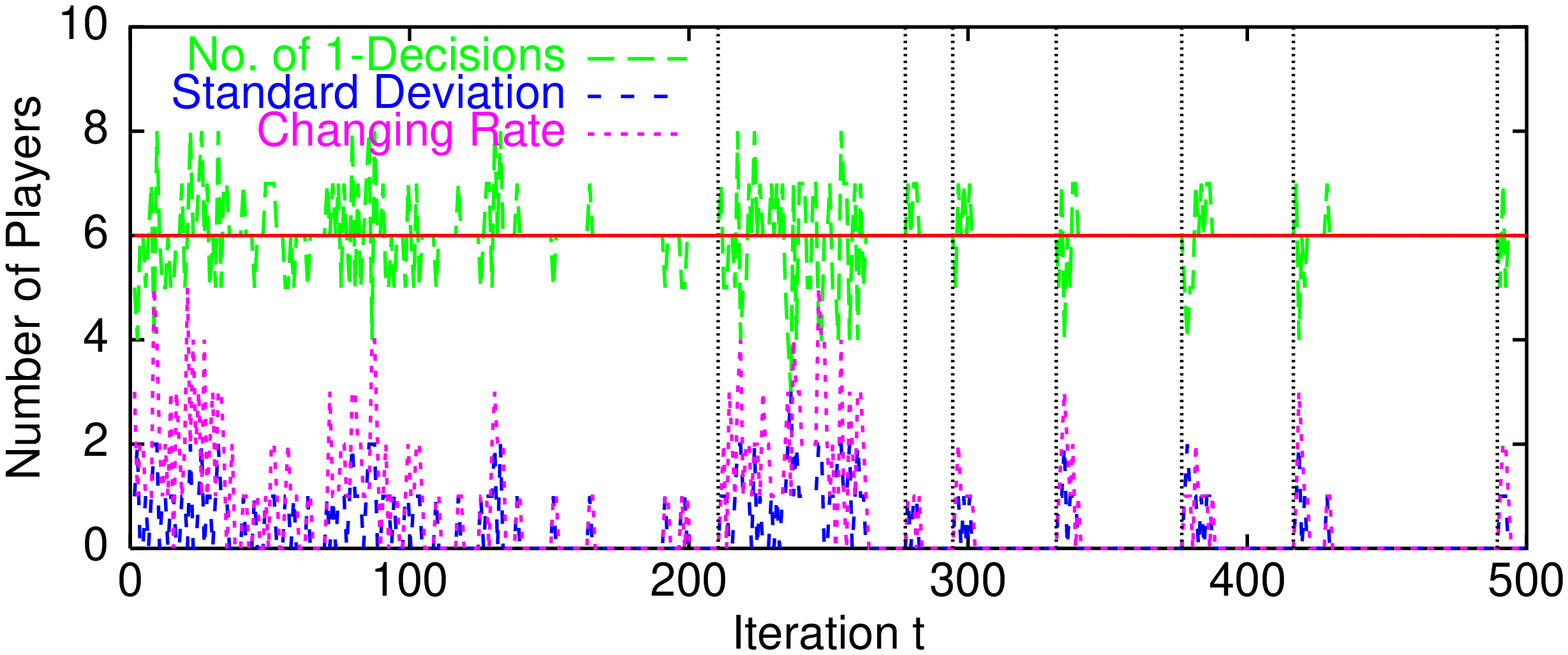}}
\put(-0.3,17.9){\footnotesize\sf (a)} 
\put(-0.3,11.8){\footnotesize\sf (b)} 
\end{picture}
\end{center}
\caption[]{Illustration of typical results for treatment 3 (from \cite{Control,Bonn}).
{(a)} Decisions of all 9 players. Players are displayed in the order of increasing
changing rate. Although the ranking of the cumulative payoff and the changing rate
are anticorrelated, the relation is not monotonic. Note that turbulent or volatile 
periods characterized by many decision changes are usually triggered by individual changes
after quiet periods (dotted vertical lines). {(b)} 
The changing rate is often larger than
the (standard) deviation from the user equilibrium $N_1 = f_1^{\rm eq}
N = 6$ (red horizontal line), 
indicating an overreaction in the system (see also Fig.~\ref{Fig7}d).} 
\label{Fig8}
\end{figure}

The efficient distribution of road capacities among users based
on individual decisions is still a fundamental problem. As individuals normally 
have aggregate information (such as radio news) rather than complete information, one frequently observes
far-from-optimal travel times. In order to learn more about this decision behavior,
we have carried out a variety of route choice experiments \cite{Control,Bonn} (see Figs.~\ref{illu} to \ref{Fig8}).
\par
In these experiments, $N$ test persons had to repeatedly decide between 
two alternative routes  $m\in\{1,2\}$ and should try to maximize their resulting payoffs $P_m$,
which were chosen proportionally to the inverse travel times. 
If the average vehicle speed $V_m$ on route $m$ is approximated by the linear relationship \cite{Gree235} 
\begin{equation}
       V_m(N_m) = V_m^0 \left( 1 - \frac{N_m(t)}{N_m^{\rm max}} \right) \, ,
\end{equation}
the inverse travel times obey the payoff relations $P_m(N_m) = P_m^0 - P_m^1 N_m$ with
\begin{equation}
 P_m^0 = \frac{V_m^0}{L_m} \qquad \mbox{and} \qquad P_m^1 = \frac{V_m^0}{N_m^{\rm max} L_m} \, . 
\end{equation}
Herein, $V_m^0$ is the maximum velocity (speed limit) and
$N_m$ the number of drivers on route $m$, $L_m$ its length, and $N_m^{\rm max}$ its capacity,
i.e. the maximum possible number of vehicles on route $m$. For an improved approach to determine
the travel times in road networks see Ref.~\cite{traveltime}.
Note that alternative routes can reach comparable payoffs and travel times only
when the total number $N = N_1+N_2$ of vehicles is large enough to fulfil the relations
$P_1(N) < P_2(0) = P_2^0$ and $P_2(N) < P_1(0) = P_1^0$. Our route choice experiment
has focussed on this traffic regime. Furthermore, we have 
the capacity restriction $N < N_1^{\rm max} + N_2^{\rm max}$, as 
$N = N_1^{\rm max} + N_2^{\rm max}$ is connected with a complete gridlock. 
\par
The Wardrop or user  equilibrium corresponds to equal travel times and payoffs for both
alternative decisions. It is found for a fraction
\begin{equation}
f_1^{\rm eq} = \frac{N_1}{N} = \frac{P_2^1}{P_1^1+P_2^1} + 
\frac{1}{N} \, \frac{(P_1^0-P_2^0)}{(P_1^1+P_2^1)} 
\end{equation}
of persons choosing alternative 1 and agrees with
the system optimum only in the limit $N \gg 1$ of many participants.
Small groups were chosen to investigate the fluctuations in the system. 
It was found that the test groups managed to adapt
to the user equilibrium on average (see Fig.~\ref{Fig7}a). However, 
although it appears reasonable to stick to the same 
decision once the  equilibrium is reached, the standard deviation stayed at a finite level. 
This was not only observed in {\em ``treatment'' 1}, where all  players knew only their own 
(previously experienced) payoff, but also in {\em treatment 2}, where
the payoffs $P'_1(N_1)$ and $P'_2(N_2)$ for both, 1- {\em and} 2-decisions, 
were transmitted to all players (analogous to radio news). Moreover,
treatment 2 could increase the average payoffs, as the additional information allowed for a better 
adaptation without having to change decisions to explore the better performing alternative
(see Ref.~\cite{Selten} and Fig.~\ref{Fig7}).
\par
With {\em treatment 3}, we managed to reveal the mysterious persistence in the changing
behavior and to achieve more than three times higher payoff increases.
Every test person was informed about
the own payoff $P'_1(N_1)$ [or $P'_2(N_2)$] {and} the {\em potential} payoff
$P'_2(N-N_1+\epsilon N)= P'_2(N_2) - \epsilon N P_2^1$ [or $P'_1(N-N_2+\epsilon N) 
= P'_1(N_1) - \epsilon N P_1^1$] 
he or she would have obtained, if a fraction $\epsilon$ of persons 
had additionally chosen the other alternative (here: $\epsilon = 1/N = 1/9$). Consequently,
in the user equilibrium with $P'_1(N_1) = P'_2(N_2)$ every player knew that
he or she would {\em not} get {\em the same}, but a {\em reduced} payoff, 
if he or she would change the decision. That explains why treatment 3 could reach 
a better adaptation performance, 
reflected by a low standard deviation and close-to-optimal average payoffs. Moreover, even the
smallest individual cumulative payoff exceeded the highest one in treatment 1. Therefore, 
treatment 3's way of information presentation is much superior to the ones used today.
\par\unitlength0.6cm
Figure~\ref{Fig8} clears up why players changed their decision in the user equilibrium at all.
We discovered intermittent behavior, i.e. quiescent periods without changes, followed by turbulent
periods with many changes. This is reminiscent of volatility clustering in stock market indices, 
where individuals also react to aggregate information summarizing all decisions 
(the trading transactions). Single individuals seem to change their decision speculating for above-average
payoffs. In fact, although the cumulative individual payoff is anticorrelated with the average changing rate,
some individuals receive higher payoffs with larger changing rates than others, i.e. they 
profit from the overreaction in the system:  Once the system is out of
equilibrium, all individuals respond in one way or another.
Typically, there are too many decision changes. 
\par\unitlength0.6cm
In {\em treatment 4}, we have tested  the group performance under changing
environmental conditions, when the participants got the same information as in
treatment 3. That is, some of the payoff parameters were varied
in time, implying a time-dependent user equilibrium. Even without recommendations,
the group managed to adapt surprisingly well to the variable conditions, but the 
standard deviation and changing rate were approximately as high as in treatment 2
(see Fig.~\ref{Fig7}). This adaptability (the collective `group intelligence') is based on complementary
responses \cite{Selten,Bonn}. That is, if some players do not react to the changing conditions, others will
take the chance to earn additional payoff. This experimentally supports the
behavior assumed in the theory of efficient markets, but here the efficiency is limited
by overreaction.
\par
To avoid overreaction, 
in {\em treatment 5} we have recommended a number $f_1^{\rm eq}(t+1)N - N_1(t)$
of players to change their decision and the other ones to keep it. These user-specific 
recommendations helped the players to reach the 
smallest overreaction of all treatments and a very low standard deviation, although the
payoffs were changing in time as in treatment 4 (see Fig.~\ref{Fig7}). 

\section{Migration and the Law of Gravity}

Urban and regional evolution, in particular the change of street infrastructures, housing,
industrial and business quarters, has a natural feedback on the spatial distribution of people.
The temporal development of the population distribution $P(\vec{x},t)$ in space $\vec{x}$ can be described
by migration models. Pioneers in this field have been Weid\-lich and Haag \cite{WeidlHa88}.
In the following, we will discuss a variant of their master equation model \cite{WeidlHa83,Weidl91,Weidl00}
\begin{equation}
 \frac{dP(\vec{x},t)}{dt} = \! \sum_{\vec{x}'} \!\!
 \, \Big[ \!\!\!\! \underbrace{w(\vec{x}|\vec{x}',t) P(\vec{x}',t)}_{\mbox{Immigration\ into\ $\vec{x}$}} 
- \underbrace{w(\vec{x}'|\vec{x},t) P(\vec{x},t)}_{\mbox{Emigration\ out\ of\ $\vec{x}$}} \!\!\!\!\! \Big] \, .
\label{migration}
\end{equation}
According to this equation, the stream of immigrants into place $\vec{x}$ is given by the fraction
$P(\vec{x}',t)$ of the population at place $\vec{x}'$ which could potentially move to place $\vec{x}$.
The proportionality factor is the transition rate $w(\vec{x}|\vec{x}',t)$ from $\vec{x}'$ to $\vec{x}$, 
i.e. the migration probability to $\vec{x}$ per unit time, given one lives at place $\vec{x}'$ before.  
Analogously, the stream of emigrants leaving place $\vec{x}$ is proportional to the fraction
$P(\vec{x},t)$ of the population living in $\vec{x}$, and the proportionality factor is the
migration rate $w(\vec{x}'|\vec{x},t)$ from $\vec{x}$ to $\vec{x}'$. This migration rate is
proportional to the fraction of people living at place $\vec{x}'$, as the number of acquaintances 
with friends, family members, or former collegues increases linearly with the number of inhabitants of
a town $\vec{x}$. However, the migration rate decreases significantly with the distance $y = \|\vec{x}-\vec{x}'\|$
between two places $\vec{x}$ and $\vec{x}'$ (and/or with the transaction costs). Accordingly, we
assume \cite{Hel95a}
\begin{equation}
 w(\vec{x}'|\vec{x},t) P(\vec{x},t) = \nu(t) \mbox{e}^{U(\vec{x}',t)}\,\mbox{e}^{-U(\vec{x},t)}
 \frac{P(\vec{x}',t)P(\vec{x},t)}{D(\|\vec{x}-\vec{x}'\|)} \, ,
\label{gravity}
\end{equation}
where $D(y)$ is a monotonously growing function in $y$.
The parameter $\nu(t)$ reflects the mobility of the population and $U(\vec{x},t)$ the utility/attractiveness of
living at place $\vec{x}$. Note that formula (\ref{gravity}) reminds of
the gravity law in physics \cite{Zip46}, where $P(\vec{x}',t)$ and $P(\vec{x},t)$ have the meaning of the
masses of two celestial bodies $\vec{x}$ and $\vec{x}'$ and $D(y) = y^2$.  Moreover,
Eq.~(\ref{gravity}) is analogous to Eq.~(\ref{analog}) with 
$b_{\vec{x}\vec{x}'} = P(\vec{x}',t)P(\vec{x},t)/D(\|\vec{x}-\vec{x}'\|)$,
$g_{\vec{x}}(t) = \mbox{e}^{-U(\vec{x},t)}$, and $h_{\vec{x}'}(t) = \mbox{e}^{U(\vec{x}',t)}$.  
That is, the gravity law can also be
motivated by entropy considerations, but this time $P(\vec{x},t)$ is adapted rather
than $g_{\vec{x}}(t)$ and $h_{\vec{x}'}(t)$. A great advantage of Eq.~(\ref{migration})
is that it can describe disequilibrium situations such as resulting migration streams when
borders between countries are opened.

\section{Urban and Regional Evolution}

Weidlich and Haag have modeled urban and regional evolution with a
master equation approach similar to Eq.~(\ref{migration}), but with
additional birth- and death-terms describing the generation or the
removal of entities from the system. Moreover, in their simulations,
they distinguished different kinds of entities (``agents''):
inhabitants, who may migrate, streets which may be expanded,
industrial areas which may grow or decay, building blocks in which
people may live, and leisure areas such as parks. Positive and
negative feedbacks between these different entities were represented
by interaction terms in the utility functions determining the
corresponding transition rates $w$. In this way, they could describe
the empirically observed differentiation in spatial usage patterns and
interdependencies between them. Moreover, they have applied their
approach to the evolution of cities in China \cite{WeidCity}.

An advantage of models based on the master equation is their huge
flexibility. A disadvantage is that there is a large number of
coefficients that need to be defined.  Therefore, there is also a
search for models that naturally include (usually geometric)
constraints and by doing so can explain some aspects of urban dynamics
with a much smaller number of free parameters.  These models originate
from the observation that some properties of cities are fractal.  For
example, the fractal dimension of the bulk of cities seems to be
around 1.9, and the fractal dimension of the (outer) perimeter of
cities seems to be around 1.3
\cite{Frankhauser:fractales-urbaines,Batty:book}.  This implies that
one might attempt to describe city growth with local growth models.
There are both attempts to explain the fractal dimension by variations
of models originally from physics
\cite{Berlin-selforg,Makse:Batty:Stanley} and attempts to re-create
(and thus predict) the growth of real-world cities
\cite{White:CA,Rabino:urban-ca}.  Yet, not all aspects of urban growth
can be explained by entirely local rules.  For example, construction
of a large mall or a golf course certainly depends on the availability
of a similar facility \emph{somewhere} in the city.  Unfortunately,
this leads to models and simulations where everything depends on
everything, which means complicated models and long simulation times.
Yet, plausibly, this is also not correct: Although large and important
facilities have a far-reaching impact, the effect still decays with
distance.
In consequence, hierarchical models, as they are known from particle
dynamics with long range interactions, are now also transfered to
urban dynamics \cite{steen:white:urban-growth}.  

As was the case at other places in this paper, the higher level
dynamics of urban growth is not independent from lower level dynamics
such as traffic congestion.  In consequence, the ultimate challenge of
this area may be to present a uniform view that bridges all scales
from car driving via human activity planning to urban evolution.
In such a view, the models presented so far would be limiting cases
for the situation where a separation of (time) scales is possible.
There are three approaches to achieve this bridging of the scales:
\begin{itemize}

\item \emph{Parameterization:} If a separation of (time) scales is
not possible, then lower level processes need to be taken into account
at least via parameterizations, i.e.\ simplified aggregated models of
the lower level processes.  As is well known, parameterization needs
to be preceeded by scientific understanding.

\item \emph{Model coupling:} Instead of using a parameterization,
i.e.\ a macroscopic description of the lower level dynamics, a
(computational) model of the lower level dynamics could be used
directly.  Such model coupling is another big challenge, not only
for computational reasons, but also because the combined model systems
are once more dynamical systems with a behavior that is usually not
well understood.

\item \emph{Microscopic across all scales:} A further alternative is
to construct a model that is microscopic or \emph{agent-based}
across all scales, that is, at all levels all decisions are made by
synthetic versions of the respective actors, usually humans.  This is
similar to a molecular dynamics simulation in physics.  It also faces
the same challenge: It is difficult to simulate large scale systems in
spite of all the computational progress we had in recent
times~\cite{computing-iatbr03}.  Nevertheless, the first partial
systems of this approach are emerging
\cite{TRANSIMS,MATSIM,ILUTE}, and validations of partial
aspects look rather promising \cite{Esser:Nagel:iatbr00-book,ch}.  
\end{itemize}
Once such models are available and operational, they could be applied
to a rather wide range of questions, ranging from zoning via
infrastructure planning to the deliberate attraction of commercial
companies and the effects of migration.  And although such models
cannot take away the responsibility for decisions from politicians or
from society, they might help that such decisions are based on a more
informed understanding of the system than today.

\section{Summary and Outlook}

\note{physics: from td limit to md sims}

Physics is a science that has a tradition in deriving large scale
emergent phenomena from microscopic rules.  Sometimes, these
derivations can be done analytically, and in such cases the
thermodynamic limit is useful.  In other situations, the microscopic
rules are too complicated to make analytical progress, and the
computer is used.  However, even with today's computational power,
only up to about $10^9$ particles can be simulated directly.

\note{number of particles in regional science in feasible range; but
particles more complex}

The situation in traffic and regional development is similar in the sense
that most effects are also caused by the combined behavior of many
particles.  In contrast to materials science, the number of particles
in regional simulations, typically several millions, is well within the
range that is computationally accessible.  On the other hand, the
rules that each individual particle follows are considerably more
complicated.

\note{first look at different levels separately, with simple models}

In this situation, it makes sense to look at different scenarios one
by one.  The first question is to identify those areas where very
simple models already lead to good levels of understanding and
prediction.  This refers, for example, to the area of traffic flow,
where already models which just include limits on acceleration,
braking, and excluded volume generate a wide range of dynamical
phenomena that can all be identified in the real world.  Similar
models are tried in the areas of route choice, activity generation,
migration, or urban growth.

\note{then use ``behavioral'' models}

If these simple models fail, then more complicated models are in
order.  These models typically contain more behavioral parameters, and
because the behavior of humans is difficult to predict, these models
are more difficult to calibrate or validate.  The challenge is, as so
often, to keep the models as simple as possible, but no simpler.
Also, one has to keep in mind that we are interested in macroscopic,
``emergent'' phenomena such as traffic jams, or general urban
patterns, and those should be easier to predict than the behavior of
individual humans.  Nevertheless, one has to get used to validation
error bars that are much larger than in physics.

\note{multi-scale.  problematic in the past ...}

With respect to real-world prediction, models that cover a large range
of scales would be useful.  Most notably, there is a rather old dream
for integrated land use transport models, in which the location
choices of inhabitants and commercial entities lead to traffic, which
in turn triggers revised location choices,
etc.~\cite{Timmermans:iatbr}  It has, however, been difficult to put
that dream into practice~\cite{Timmermans:iatbr}, and one could argue
that the main obstacle has been that the interactions between the
scales were not well understood but rather included in an ad-hoc
manner.  Simultaneously, computing power in the past has not really
been sufficient for sound solutions.

\note{... but looking better for the future}

However, in our view the situation is now changing.  Based on the
combination of computational, analytical, and experimental work, there
has been considerable progress with respect to understanding aspects
at the different levels, leading to better and faster submodels in the
future.  At the same time, it is now possible to build computational
models that bridge large ranges of scales while remaining microscopic,
which, although computationally demanding, again solves a large
variety of conceptual problems.

\note{Data}

Finally, one should also mention that data availability in the field
of geo-science has made a quantum jump forward.  Satellite-derived
high resolution spatial data is now available for nearly everywhere;
decent census data is available in nearly every industrialized
country; vector data for the infrastructure is increasingly becoming
available; and a whole range of telematics devices is capable of
collecting important data about spatial behavior, although aspects of
privacy need to be valued against scientific interest.  This also
means that the traditional approach of physics, where theory,
computation, and real-world measurement enhance each other, can now be
applied to traffic and regional systems.

Overall, it is our impression that the new computing and data
collection technologies have had a huge impact in the area of traffic
and regional systems, and that both the computational and the
theoretical work are struggling to keep up with the new possibilities.
This, combined with the obvious societal relevance of the area, makes
this a very exciting field. 

\section{Authors' Biographies}

Dirk Helbing (*19/Jan/1965) is the Managing Director of the Institute
for Economics and Traffic at Dresden University of Technology.
Originally, he studied Physics and Mathematics in G\"ottingen,
Germany, but soon he got fascinated in interdisciplinary problems.
Therefore, his master thesis dealt with physical models of pedestrian
dynamics, while his Ph.D. thesis at the University of Stuttgart
focussed on modeling interactive decisions and behaviors with methods
from statistical physics and the theory of complex systems. After his
habilitation on the physics of traffic flows, he received a Heisenberg
scholarship and worked at Xerox PARC in Silicon Valley, the Weizmann
Institute in Israel, and the Collegium Budapest---Institute for
Advanced Study in Hungary.  He gets excited when physics meets traffic
or social science, economics, or biology, and if the results are
potentially relevant for everyday life.

Kai Nagel (*17/Sep/1965) is Professor for Transport Systems Analysis
and Transport Telematics at the Institute of Land and Sea Transport
Systems at the Technical University Berlin in Germany.  He studied
physics and meteorology at the University of Cologne and the
University of Paris, with one master's thesis in the area of cellular
automata models for cloud formation and another one in the area of
large scale climate simulations.  His Ph.D., in computer science at
the University of Cologne, was about cellular automata models for
large scale traffic simulations.  He then was postdoc, staff member,
and research team leader at Los Alamos National Laboratory, working on
the TRANSIMS (Transportation ANalysis and SIMulation System) project.
In 1999--2004, he was assistant professor for computer science at ETH
Zurich in Switzerland.  His interests are in large scale simulation
and in the simulation and modeling of socio-economic systems.


\begin{references} 

\bibitem{Review}
Helbing, D., 2001, Traffic and related self-driven many-particle systems.  
{\em Reviews of Modern Physics}, {\bf 73}(4), 1067--1141.

\bibitem{Ency} 
Helbing, D., 2004, Traffic. In {\em Encyklopedia of Nonlinear Science}, (London: Taylor \& Francis).

\bibitem{LiWi}
Lighthill, M. J., and Whitham, G. B., 1955, On kinematic waves: {II. A} theory of traffic on long crowded roads. 
{\em Proc. Roy. Soc. London, Ser. A}, {\bf 229}, 317--345.

\bibitem{KerKon94} 
Kerner, B. S., and Konh\"auser, P., 1994, Structure and parameters of clusters in traffic flow. 
{\em Phys. Rev. E}, {\bf 50}, 54--83. 

\bibitem{KerKoSchi95}
Kerner, B. S., Konh\"auser, P., and Schilke, M., 1995, Deterministic spontaneous appearance of traffic jams in slightly inhomogeneous traffic flow. 
{\em Physical Review E}, {\bf 51}, R6243--R6246.

\bibitem{KerKoSc96}
Kerner, B. S., Konh\"auser, P., and Schilke, M., 1996, `Dipole-layer' effect in dense traffic flow. 
{\em Physics Letters A}, {\bf 215}, 45--56.

\bibitem{Ker97}
Kerner, B. S., Klenov, S. L., and Konh\"auser, P., 1997, Asymptotic theory of traffic jams. 
{\em Phys. Rev. E}, {\bf 56}, 4200--4216.

\bibitem{Krauss} 
Krau{\ss}, S., 1998, {\em Microscopic Modeling of Traffic Flow: Investigation of Collision
Free Vehicle Dynamics} (Cologne: DLR---Deutsches Zentrum f\"ur Luft- und Raumfahrt e.V.).

\bibitem{KerReh96a} 
Kerner, B. S., and Rehborn, H., 1996, Experimental features and characteristics of traffic jams. 
{\em Phys. Rev. E}, {\bf 53}, R1297--R1300.

\bibitem{HelHenTre99} 
Helbing, D., Hennecke, A., and Treiber, M., 1999, {Phase diagram of traffic states in the presence of inhomogeneities}. 
{\em Phys. Rev. Lett.}, {\bf 82}, 4360--4363.

\bibitem{HelHeShTr01a}
Helbing, D., Hennecke, A., Shvetsov, V., and Treiber, M., 2001, Macroscopic traffic simulation based on a gas-kinetic, non-local traffic model. 
{\em Transportation Research B}, {\bf 35}, 183--211.

\bibitem{HelHeShTr01b}
Helbing, D., Hennecke, A., Shvetsov, V., and Treiber, M., 2002, Micro- and macrosimulation of freeway traffic. 
{\em Mathematical and Computer Modelling}, {\bf 35}, 517--547.

\bibitem{PriAnd}
Prigogine, I., and Andrews, F. C., 1960, A {B}oltzmann-like approach for traffic flow.
{\em Opns. Res.}, {\bf 8}, 789--797.

\bibitem{Prig}
Prigogine, I., 1961, A {B}oltzmann-like approach to the statistical theory of traffic flow. 
In {\em Theory of Traffic Flow}, edited by R. Herman (Amsterdam: Elsevier), pp. 158--164.

\bibitem{PriHer}
Prigogine I., and Herman, R., 1971, {\em Kinetic Theory of Vehicular Traffic}  
(New York: Elsevier). 

\bibitem{HeTre98}
Helbing D., and Treiber, M., 1998, {Gas-kinetic-based traffic model explaining observed hysteretic phase transition}. 
{\em Phys. Rev. Lett.}, {\bf 81}, 3042--3045.

\bibitem{ShvHe99}
Shvetsov, V., and Helbing, D., 1999, {Macroscopic dynamics of multilane traffic}. 
{\em Phys. Rev. E}, {\bf 59}, 6328--6339.

\bibitem{Payne1}
Payne, H. J., 1971, {Models of freeway traffic and control}. 
In {\em Mathematical Models of Public Systems}, edited by G.~A. Bekey (La Jolla, CA: Simulation Council), Vol. 1, pp. 51--61.

\bibitem{Payne2}
Payne, H. J., 1979, A critical review of a macroscopic freeway model. 
In {\em Research Directions in Computer Control of Urban Traffic Systems}, 
edited by W.~S. Levine, E. Lieberman, and J.~J. Fearnsides (New York: American Society of Civil Engineers), pp. 251--265.

\bibitem{KerKon93} 
Kerner, B. S., and Konh{\"a}user, P., 1993, {Cluster effect in initially homogeneous traffic flow}. 
{\em Phys. Rev. E}, {\bf 48}, R2335--R2338.

\bibitem{Kuehne}
K{\"u}hne, R. D., 1984, {Macroscopic freeway model for dense traffic---{S}top-start waves and incident detection}. 
{\em Proceedings of the 9th International Symposium on Transportation and Traffic Theory}, edited by I. Volmuller, and R. Hamerslag 
(Utrecht: VNU Science), pp. 21--42.

\bibitem[Kerner and Konh\"auser(1994)]{Kerner:Konh:large:amplitude}
Kerner, B. S., and Konh\"auser, P., 1994, Structure and parameters of clusters in traffic flow. 
{\em Phys. Rev. E}, {\bf 50}, 54--83.

\bibitem{Lee3LeKi99}
Lee, H. Y., Lee, H.-W., and Kim, D, 1999, Dynamic states of a continuum traffic equation with on-ramp. 
{\em Phys. Rev. E}, {\bf 59}, 5101--5111.

\bibitem{Lee3LeKi00}
Lee, H. Y., Lee, H.-W., and Kim, D, 2000, Phase diagram of congested traffic flow: An empirical study. 
{\em Phys. Rev. E}, {\bf 62}, 4737--4741. 

\bibitem{inTGF01} 
Helbing, D., {\em et al.}, 2003, Critical discussion of ``synchronized flow'', pedestrian evacuation, and optimal production. 
In {\em Traffic and Granular Flow '01}, edited by M. Fukui, {\em et al.}, (Berlin: Springer), pp. 511--530.

\bibitem{TS}
Sch\"onhof, M., and Helbing, D., 2004, Empirical features of congested traffic states and their implications for traffic modeling. 
In preparation for {\em Transportation Science}.

\bibitem{TreHeHe00} 
Treiber, M., Hennecke, A., and Helbing, D., 2000, Congested traffic states in empirical observations and microscopic simulations. 
{\em Phys. Rev. E}, {\bf 62}, 1805--1824. 

\bibitem{Koshi}
Koshi, M., Iwasaki, M., and Ohkura, I., 1983, Some findings and an overview on vehicular flow characteristics. 
{\em Proceedings of the 8th International Symposium on Transportation and Traffic Flow Theory}, edited by V. F. Hurdle, E. Hauer, and G.~N. Stewart (Toronto, Ontario: University of Toronto), pp. 403--426. 

\bibitem{pinch}
Kerner, B. S., 1998, Experimental features of self-organization in traffic flow. 
{\em Phys. Rev. Lett.}, {\bf 81}, 3797--3800.

\bibitem{KerReh96b} 
Kerner, B. S., and Rehborn, H, 1996, Experimental properties of complexity in traffic flow. 
{\em Phys. Rev. E}, {\bf 53}, R4275--R4278.

\bibitem{critic1} 
Kerner, B. S., 1998, A theory of congested traffic flow. 
{\em Proceedings of the 3rd International Symposium on Highway Capacity}, edited by R. Rysgaard (Denmark: Road Directorate), Vol. 2, pp. 621--642. 

\bibitem{critic2}
Kerner, B. S., 2000, Phase transitions in traffic flow. 
In {\em Traffic and Granular Flow '99}, edited by D. Helbing, H. J. Herrmann, M. Schreckenberg, and D. E. Wolf (Berlin: Springer), pp. 253--284.

\bibitem{critic3}
Kerner, B. S., 2000, Theory of breakdown phenomenon at highway bottlenecks. 
{\em Transpn. Res. Rec.}, {\bf 1710}, 136--144. 

\bibitem{Scattering}
Treiber, M., and Helbing, D., 1999, Macroscopic simulation of widely scattered synchronized traffic states. 
{\em J. Phys. A: Math. Gen.}, {\bf 32}, L17--L23.

\bibitem{Kats} 
Nishinari, K., Treiber, M., and Helbing, D., 2003, Interpreting the wide scattering of synchronized traffic data by time gap statistics. 
{\em Physical Review E}, {\bf 68}, 067101.

\bibitem{Gazis}
Gazis, D. C., Herman, R., and Rothery, R. W., 1961, {Nonlinear follow the leader models of traffic flow}. 
{\em Opns. Res.}, {\bf 9}, 545--567. 

\bibitem{Bando1}
Bando, M., Hasebe, K., Nakayama, A., Shibata, A., and Sugiyama, Y., 1994, Structure stability of congestion in traffic dynamics. 
{\em Jpn. J. Industr. Appl. Math.}, {\bf 11}, 203--223. 

\bibitem{Bando2}
Bando, M., Hasebe, K., Nakayama, A., Shibata, A., and Sugiyama, Y., 1995, {Dynamical model of traffic congestion and numerical simulation}. 
{\em Phys. Rev. E}, {\bf 51}, 1035--1042. 

\bibitem{Mehta}
Mehta, K., 1991, {\it Random Matrices} (New York: Academic Press).

\bibitem{fpg} 
Helbing, D., and Treiber, M., 2003, Fokker-Planck equation approach to vehicle statistics. 
Preprint http://arXiv.org/abs/cond-mat/0307219.

\bibitem{milan}
Krbalek, M. and Helbing, D., 2004, Determination of interaction potentials in freeway traffic from steady-state statistics. 
{\em Physica A}, {\bf 333}, 370--378. 

\bibitem{Neubert}
Neubert, L.,  Santen, L., Schadschneider, A., Schreckenberg, M., 1999, 
{\it Physical Review E}, {\bf 60}, {6480}.

\bibitem{MayKe67}
Ma{y, Jr.}, A.~D., and Keller, H. E. M., 1967, Non-integer car-following models. 
{\em Highway Res. Rec.}, {\bf 199}, 19--32. 

\bibitem{PhysJ}
Helbing, D., and Nagel, K., 2003, Verkehrsdynamik und urbane Systeme. 
{\em Physik Journal}, {\bf 2}(5), 35--41. 

\bibitem{May} 
May, A. D., 1990, {\em Traffic Flow Fundamentals} 
(Englewood Cliffs, NJ: Prentice Hall).

\bibitem{Hoog} 
Hoogendoorn, S. P., and Bovy, P. H. L., 1998, 
{\em Motorway Traffic Flow Analysis}, edited by P. H. L. Bovy (Delft: University of Technology), pp. 71.

\bibitem{Gazis:queue}
Gazis, D. C., 1974, Modeling and optimal control of congested transportation systems. 
{\em Networks}, {\bf 4}, 113--124.

\bibitem{Gawron:queue}
Gawron, C., 1998, An iterative algorithm to determine the dynamic user equilibrium in a traffic simulation model. 
{\em International Journal of Modern Physics C}, {\bf 9}, 393--407.

\bibitem[Cetin and Nagel(2003)]{queue}
Cetin, N., and Nagel, K., 2003, A large-scale agent-based traffic microsimulation based on queue model. 
{\em Proceedings of Swiss Transport Research Conference (STRC)} (Monte Verita, CH).

\bibitem[Nagel and Raney(2004)]{Nagel:Raney:w-schmid-spatial-planning} 
Nagel, K., and Raney, B., 2004, Complex systems applications for transportation planning. In {\em The Real and Virtual Worlds of Spatial Planning}, edited by M.~Koll-Schretzenmayr, M.~Keiner, and G.~Nussbaumer (Berlin: Springer-Verlag), chapter 16.

\bibitem{Lohse1}
Schnabel, W., and Lohse, D., 1997, {\em Grundlagen der Stra{\ss}enverkehrstechnik und der Verkehrsplanung}, Band 2  
(Berlin: Verlag f\"ur Bauwesen).

\bibitem{Lohse2}
Lohse, D., Teichert, H., Dugge, B., and Bachner, G., 1997, {Ermittlung von Verkehrsstr\"omen mit $n$-linearen Gleichungssystemen~-- Verkehrsnachfragemodellierung}. 
In {\em Schriftenreihe des Instituts f\"ur Verkehrsplanung und Stra{\ss}enverkehr, TU Dresden}, Heft 5/1997.

\bibitem{Lohse3}
Lohse, D., and Schneider, R., 1997, {Vergleichende Untersuchungen der aggregierten und disaggregierten Verkehrsplanungsmodelle}. 
In {\em Schriftenreihe des Instituts f\"ur Verkehrsplanung und Stra{\ss}enverkehr, TU Dresden}, Heft 3/1997.

\bibitem{eins} 
DETR/DTLR Data sn2852, sn2853, sn2854, sn2855, sn3288, sn4108, 1998, 2000, (Essex: The Data Archive).

\bibitem{drei} 
Hettinger, T., 1989, Physiologische Leistungsgrundlagen.  
In {\em Handbuch der Ergonomie}, 2nd ed., edited by H. Schmidke (Munich: Hanser), Vol. 2.

\bibitem{vier1} 
Spitzer, H., Hettinger, T., and Kaminsky, G., 1982, {\em Tafeln f\"{u}r den Energieumsatz bei k\"{o}rperlicher Arbeit}, 6th ed. (Berlin: Beuth).

\bibitem{vier2} 
Goodwin, P. B., 1976, Human effort and the value of travel time. 
{\em Journal of Transport Economics and Policy}, {\bf 10}, 3--15. 

\bibitem{vier3}
Rowland, T. W., 1998, The biological basis of physical activity. 
{\em Medicine \& Science in Sports \& Exercise}, 392-399.

\bibitem{robert} 
K\"olbl, R., and Helbing, D., 2003, Energy laws in human travel behaviour. 
{\em New Journal of Physics}, {\bf 5}, 48.1.--48.12. 

\bibitem{f1} 
Zahavi, Y., and Talvitie, A., 1980, Regularities in travel time and money expenditure. 
{\em Transpn. Res. Rec.}, {\bf 750}, 13--19. 

\bibitem{f2} 
Zahavi, Y., and Ryan, J., 1980, Stability of travel components over time. 
{\em Transpn. Res. Rec.}, {\bf 750}, 19--26.

\bibitem{f3} 
Roth, G. J., and Zahavi, Y., 1981, Travel time ``budgets'' in developing countries. 
{\em Transpn. Res. A}, {\bf 15}, 87--95. 

\bibitem{f4} 
Supernak, J., and Zahavi, Y., 1982, Travel-time budget: A critique (discussion). 
{\em Transpn. Res. Rec.}, {\bf 28}(879), 15--28. 

\bibitem{f5} 
Tanner, J. C., 1981, Expenditure of time and money on travel. 
{\em Transpn. Res. A}, {\bf 15}, 25--38.

\bibitem{f6} 
Schafer, A., 1998, The global demand for motorized mobility. 
{\em Transpn. Res. A}, {\bf 32}, 455--477.

\bibitem{f7} 
Schafer, A., 2000, Regularities in travel demand: An international perspective. 
{\em Journal of Transportation and Statistics}, {\bf 3}(3), 1--31. 

\bibitem{f8} 
Goodwin, P. B., 1981, The usefulness of travel budgets. 
{\em Transpn. Res. A}, {\bf 15}, 97--106. 

\bibitem{MNL}  
Domencich, T. A., and McFadden, D., 1975, {\em Urban Travel Demand. A Behavioral Analysis}, pp.~61--69; Ort\'{u}zar, J. d. D., and Willumsen, L. G., 1994, {\em Modelling Transport}, 2nd ed. (New York: Wiley).

\bibitem{s1} 
Wilson, A. G., 1967, A statistical theory of spatial distribution modes. 
{\em Transpn. Res. A}, {\bf 1}, 253--269.

\bibitem{s2}
Wilson, A. G., 1970, {\em Entropy in urban and regional modelling} (London: Pion).

\bibitem{s3} 
Wilson, A. G., 1998, Land-use/Transport Interaction Models. Past and Future.  
{\em Journal of Transport Economics and Policy}, {\bf 32}, 3--26.

\bibitem{s4}
Ben-Akiva, M., and Lerman, S. R., 1997, {\em Discrete Choice Analysis: Theory and Application to Travel Demand} (Cambridge, MA: MIT Press).

\bibitem{s5} 
Ben-Akiva, M., McFadden, D. M. \emph{et al.}, 1999, {Extended framework for modeling choice behavior}. 
{\em Marketing Lett.}, {\bf 10}, 187--203.

\bibitem{iatbr00-book}
Hensher, D., and King, J., 2001. In {\em The Leading Edge of Travel Behavior Research}, edited by D. Hensher, and J. King (Oxford: Pergamon).

\bibitem{Ben-Akiva:book}
Ben-Akiva, M., and Lerman, S. R., 1985, {\em Discrete choice analysis} (Cambridge, MA: The MIT Press).

\bibitem[{BIOGEME www page}(accessed 2004)]{biogeme} 
{BIOGEME www page}. www.roso.epfl/biogeme, accessed 2004.

\bibitem{Bowman:thesis}
Bowman, J. L., 1998, The day activity schedule approach to travel demand analysis, PhD thesis (Cambridge, MA: Massachusetts Institute of Technology).

\bibitem{Bowman:etc:Portland:acts}
Bowman, J. L.,  Bradley, M., Shiftan, Y., Lawton, T. K., and Ben-Akiva, M., 1999, {\em Demonstration of an activity-based model for {P}ortland}, Vol. 3 (Oxforf: Elsevier).

\bibitem{Kitamura:CO2}
Kitamura, R.,  Fujii, S., Kikuchi, A., and Yamamoto, T., 1997, An application of a micro-simulator of daily travel and dynamic network flow to evaluate the effectiveness of selected tdm measures for {CO$_2$} emissions reduction.

\bibitem{q4acts}
Charypar, D., Graf, P., and Nagel, K., 2004, Q-learning for flexible learning of daily activity plans. 
{\em Proceedings of Swiss Transport Research Conference (STRC)} (Monte Verita, CH), see: www.strc.ch.

\bibitem{ALBATROSS}
Arentze, T., Hofman, F., {van Mourik}, H., and Timmermans, H., 2000, {ALBATROSS}: A multi-agent rule-based model of activity pattern decisions, Paper~22, Transportation Research Board Annual Meeting, Washington, D.C..

\bibitem{ga4acts}
Charypar, D., and Nagel, K., 2003, Generating complete all-day activity plans with genetic algorithms. 
{\em Proceedings of the meeting of the International Association for Travel Behavior Research (IATBR)} (Lucerne, Switzerland), see http://www.ivt.baum.ethz.ch/allgemein/iatbr2003.html.

\bibitem{Unger:phd}
Unger, H., 2002, {\em Modellierung des Verhaltens autonomer Verkehrsteilnehmer in einer variablen staedtischen Umgebung}, PhD thesis, TU Berlin, 2002.

\bibitem{arentze:mental-map:iatbr03}
Arentze, T., and Timmermans, H. J. P., 2003, Representing mental maps and cognitive learning in micro-simulation models of activity-travel choice dynamics. 
{\em Proceedings of the meeting of the International Association for Travel Behavior Research (IATBR)} (Lucerne, Switzerland), see http://www.ivt.baum.ethz.ch/allgemein/iatbr2003.html.

\bibitem[Kistler(2004)]{dkistler:masters}
Kistler, D., 2004, Mental maps for mobility simulations of agents, Master's thesis, ETH Zurich.

\bibitem[Hofbauer and Sigmund(1998)]{Hofb:Sigm:book} 
Hofbauer, J., and Sigmund, K., 1998, {\em Evolutionary games and replicator dynamics} (Cambridge: University Press).

\bibitem{Sheffi:book}
Sheffi, Y., 1985, {\em Urban transportation networks: Equilibrium analysis with mathematical programming methods} (Englewood Cliffs: Prentice-Hall).

\bibitem{Cascetta:Cantarella:day2day}
Cascetta, E., and Cantarella, C., 1991, A day-to-day and within day dynamic stochastic assignment model. 
{\em Transportation Research A}, {\bf 25}, 277--291.

\bibitem{Palmer:broken:ergodicity}
Palmer, R., 1989, Broken ergodicity. 
In {\em Lectures in the Sciences of Complexity}, edited by D.~L. Stein (Redwood City, CA: Addison-Wesley), Vol. I of {\em Santa Fe Institute Studies in the Sciences of Complexity}, pp. 275--300.

\bibitem{Zuylen:Taale:tgf03}
van Zuylen, H. J., and Taale, H., in press.
In {\em Traffic and granular flow~'03}, edited by P. Bovy et al.

\bibitem{Nagel:etc:sqrt-utl}
Nagel, K., Strauss, M., and Shubik, M., in press, {\em Physica A}.

\bibitem{Wardrop} 
Wardrop, J. G., 1952, Some theoretical aspects of road traffic research.  
{\em Proceedings of the Institution of Civil Engineers II}, Vol. 1, pp. 325--378.

\bibitem{Control} 
Helbing, D., Sch\"onhof, M., and Kern, D., 2002,  Volatile decision dynamics: Experiments, stochastic description, intermittency control, and traffic optimization. 
{\em New Journal of Physics}, {\bf 4}, 33.1.--33.16.

\bibitem{Bonn}
Helbing, D., 2004, Dynamic decision behavior and optimal guidance through information services: Models and experiments.  
In {\em Human Behaviour and Traffic Networks}, edited by M. Schreckenberg, and R. Selten (Berlin: Springer).

\bibitem{Gree235}
Greenshields, B.~D., 1935, A study of traffic capacity. 
{\em Proceedings of the Highway Research Board}, (Washington, D. C.: Highway Research Board), Vol. 14, pp. 448--477.

\bibitem{traveltime} 
Helbing, D., 2003, A section-based queueing-theoretical traffic model for congestion and travel time analysis in networks. 
{\em J. Phys. A: Math. Gen.}, {\bf 36}, L593--598. 

\bibitem{Selten} 
Schreckenberg, M., Selten, R., Chamura, T., Pitz, T., and Wahle, J., 2001, Experiments on day-to-day route choice. 
www.trafficforum.org, e-print 01080701.

\bibitem{WeidlHa88} 
Weidlich, W., and Haag, G., 1988, {\em Interregional Migration} (Berlin: Springer).

\bibitem{WeidlHa83}
Weidlich, W., and Haag, G., 1983, {\em Concepts and Models of a Quantitative Sociology. The Dynamics of Interacting Populations} (Berlin: Springer).

\bibitem{Weidl91}
Weidlich, W., 1991, Physics and social science---{T}he approach of synergetics. 
{\em Physics Reports}, {\bf 204}, 1--163. 

\bibitem{Weidl00} 
Weidlich, W., 2000, {\em Sociodynamics. A Systematic Approach to Mathematical Modelling in the Social Sciences} (Amsterdam: Harwood Academic).

\bibitem{Hel95a}
Helbing, D., 1995, {\em Quantitative Sociodynamics. Stochastic Methods and Models of Social Interaction Processes} (Dordrecht: Kluwer Academic).

\bibitem{Zip46} 
Zipf, G.~K., 1946, The {$P1P2/D$} hypothesis on the intercity movement of persons. 
{\em Amer. Sociol. Rev.}, {\bf 11}, 677--686. 

\bibitem{WeidCity}
Weidlich, W., and Haag, G., 1999, {\em An Integrated Model of Transport and Urban Evolution} (Berlin: Springer).

\bibitem{Frankhauser:fractales-urbaines} 
Frankhauser, P., 1994, {\em La fractalit{\'e} des structrures [sic] urbaines} (Paris: Anthropos).

\bibitem{Batty:book} 
Batty, M., and Longley, P., 1994, {\em Fractal Cities: {A} Geometry of Form and Function} (Academic Press).

\bibitem{Berlin-selforg} 
Schweitzer, F., 1997, {\em Self-organization of complex structures: {F}rom individual to collective dynamics} (London: Gordon and Breach).

\bibitem{Makse:Batty:Stanley}
Makse, H. A., {Andrade, Jr.}, J. S., Batty, M., Havlin, S., and Stanley, H. E., 1998, Modeling urban growth patterns with correlated percolation. 
{\em Physical Review E}, {\bf 58}, 7054--7062.

\bibitem{White:CA}
White, R., Engelen, G., and Uljee, I., 1997, The use of constrained cellular automata for high-resolution modeling of urban landscape dynamics. 
{\em Environment and Planning B}, {\bf 24}, 323--343. 

\bibitem{Rabino:urban-ca}
Rabino, G., and Laghi, A., 2002, Identification of cellular automata: theoretical remarks. 
{\em Proceedings of the Annual Conference of The European Regional Science Association (ERSA)} (Dortmund), see: www.ersa.org.

\bibitem{steen:white:urban-growth}
Andersson, C., Lindgren, K., Rasmussen, S., and White, R., 2002, Urban growth simulation from `first principles'. 
{\em Physical Review E}, {\bf 66}, 026204.

\bibitem[Nagel and Marchal(2003)]{computing-iatbr03}
Nagel, K., and Marchal, F., 2003, Computational methods for multi-agent simulations of travel behavior. {\em Proceedings of the meeting of the International Association for Travel Behavior Research (IATBR)} (Lucerne), see http://www.ivt.baum.ethz.ch/allgemein/iatbr2003.html.

\bibitem[{MATSIM www page}(accessed 2004)]{MATSIM}
{MATSIM www page}. {M}ulti{A}gent {T}ransportation {SIM}ulation. see www.matsim.org, accessed 2004.

\bibitem[Salvini and Miller(2003)]{ILUTE}
Salvini, P. A., and Miller, E. J., 2003, {ILUTE}: {A}n operational prototype of a comprehensive microsimulation model of urban systems. {\em Proceedings of the meeting of the International Association for Travel Behavior Research (IATBR)} (Lucerne), see http://www.ivt.baum.ethz.ch/allgemein/iatbr2003.html.

\bibitem[{TRANSIMS www page}(accessed 2004)]{TRANSIMS}
{TRANSIMS www page}. {TR}ansportation {AN}alysis and {SIM}ulation {S}ystem. transims.tsasa.lanl.gov, accessed 2004, {L}os Alamos National Laboratory, Los Alamos, NM.

\bibitem[Esser and Nagel(2001)]{Esser:Nagel:iatbr00-book}
Esser, J., and Nagel, K., 2001, Iterative demand generation for transportation simulations. In \cite{iatbr00-book}, pp. 689--709.

\bibitem[Raney et~al.(2003)Raney, Cetin, V{\"o}llmy, Vrtic, Axhausen, and
  Nagel]{ch} 
Raney, B., Cetin, N., V{\"o}llmy, A., Vrtic, M., Axhausen, K., and Nagel, K., 2003, An agent-based microsimulation model of {S}wiss travel: First results. 
{\em Networks and Spatial Economics}, {\bf 3}, 23--41.

\bibitem[Timmermans(2003)]{Timmermans:iatbr}
Timmermans, H. J. P., 2003, The saga of integrated land use-transport modeling: How many more dreams before we wake up? 
{\em Proceedings of the meeting of the International Association for Travel Behavior Research (IATBR)} (Lucerne), see http://www.ivt.baum.ethz.ch/allgemein/iatbr2003.html.
\end{references}
\end{document}